\documentclass[sigconf]{acmart}
\usepackage{booktabs} %
\usepackage{subfigure}
\usepackage{multirow}
\usepackage{epstopdf}
\usepackage[strings]{underscore}
\usepackage[english]{babel}
\usepackage{color, colortbl}
\usepackage{balance}
\setcopyright{rightsretained}
\usepackage{setspace}
\usepackage{amsmath}
\DeclareMathOperator*{\argmax}{argmax} %

\def \system {Hapi}

\def \sys {\textit{\system}}
\definecolor{pistachio}{rgb}{0.58, 0.77, 0.45}
	\definecolor{wolf}{rgb}{0.86, 0.84, 0.82}
\definecolor{green}{rgb}{0.26, 0.7, 0.68}
\definecolor{aliceblue}{rgb}{0.94, 0.97, 1.0}
	\definecolor{champagne}{rgb}{0.97, 0.91, 0.81}
\definecolor{almond}{rgb}{0.94, 0.87, 0.8}
\definecolor{white}{rgb}{0.95, 0.95, 0.96}
	\definecolor{azure}{rgb}{0.94, 1.0, 1.0}	
\definecolor{blue}{rgb}{0.45, 0.66, 0.76}
\begin{document}
\title[\system]{\system: A Robust Pseudo-3D Calibration-Free WiFi-based Indoor Localization System}

\author{Heba Aly}
\affiliation{%
  \department{Department of Computer Science}
    \institution{University of Maryland}
  \city{College Park}
  \state{MD}
  \postcode{20742}
  \country{USA}
  }
  \email{heba@cs.umd.edu}
  \author{Ashok Agrawala}
\affiliation{%
  \department{Department of Computer Science}
  \institution{University of Maryland}
  \city{College Park}
  \state{MD}
  \postcode{20742}
  \country{USA}
  }
  \email{agrawala@cs.umd.edu}

\renewcommand{\shortauthors}{Heba Aly et al.}

\begin{abstract}

In this paper, we present \sys{}, a novel system that uses off-the-shelf standard WiFi to provide pseudo-3D indoor localization---it estimates the user's floor and her 2D location on that floor. \sys{} is calibration-free, \emph{only requiring the building's floorplans and its WiFi APs' installation location for deployment}. Our analysis shows that while a user can hear APs from nearby floors as well as her floor, she will typically only receive signals from spatially closer APs in distant floors, as compared to APs in her floor. This is due to signal attenuation by floors/ceilings along with the 3D distance between the APs and the user. \sys{} leverages this observation to achieve \emph{accurate and robust} location estimates.

A deep-learning based method is proposed to identify the user's floor. 
Then, the identified floor along with the user's visible APs from all floors are used to estimate her 2D location through a novel RSS-Rank Gaussian-based method.
Additionally, we present a regression based method to predict \sys{}'s location estimates' quality and employ it within a Kalman Filter to further refine the accuracy. 
 Our evaluation results, from deployment on various android devices \textbf{over 6 months with 13 subjects in 5 different up to 9 floors multistory buildings}, show that \sys{} can identify the user's exact floor up to 95.2\% of the time and her 2D location with a median accuracy of 3.5m, achieving 52.1\% and 76.0\% improvement over related calibration-free state-of-the-art systems respectively.

\end{abstract}



\copyrightyear{2018} 
\acmYear{2018} 
\setcopyright{acmcopyright}
\acmConference[MobiQuitous '18]{EAI International Conference on Mobile and Ubiquitous Systems: Computing, Networking and Services}{November 5--7, 2018}{New York, NY, USA}
\acmBooktitle{EAI International Conference on Mobile and Ubiquitous Systems: Computing, Networking and Services (MobiQuitous '18), November 5--7, 2018, New York, NY, USA}
\acmPrice{15.00}
\acmDOI{10.1145/3286978.3286980}
\acmISBN{978-1-4503-6093-7/18/11}

\maketitle

\section{Introduction}

Location is one of the most valuable user-context information, with a rapidly growing number of Location-based Services (LBSs) becoming an integral part of our daily life~\cite{muratasmartphone,aly2016robust,aly2015lanequest,aly2017towards,aly2014map++,aly2017automatic}. Nevertheless, we are still missing a ubiquitous indoor localization system~\cite{youssef2015towards}. 
The world-wide ubiquity of WiFi networks, attracted lots of attention from researchers to utilize the already available WiFi infrastructure for identifying the user's indoor location~\cite{wang2012no,bhargava2015locus,youssef2005horus,liu2016selective,elbakly2016robust,abdelnasser2016semanticslam,he2018slac}. Yet, currently available systems entail a tedious calibration/training phase for each deployment building/floor, require additional sensors, have a coarse-grained accuracy and/or sensitive to device/environment heterogeneity, among others. Another major additional limitation for the vast majority of available systems, e.g.~\cite{wang2012no,youssef2005horus,liu2016selective,elbakly2016robust,abdelnasser2016semanticslam,he2018slac}, is focusing on a single-floor area of interest %
which reduces the usefulness of their estimated location in realistic %
 multistory buildings scenarios~\cite{bhargava2015locus}.

\begin{figure}[!t]
\center
\includegraphics[width=0.75\linewidth,height=3.2cm]{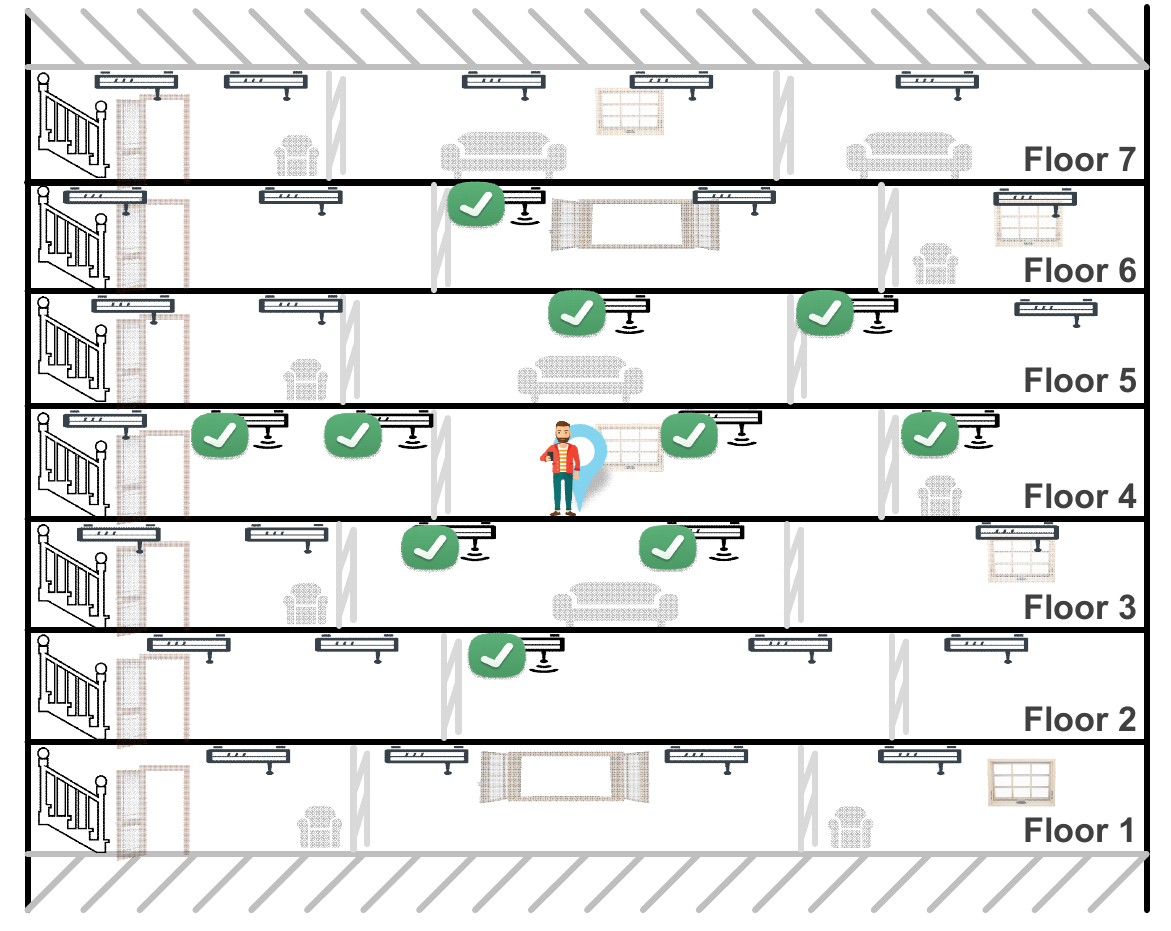}
\vspace*{-2mm}
\caption{A user standing at floor 4 has visible APs (marked with a tick) from nearby floors along with her actual one (from floor 2 to 6). However, the user is hearing more distant APs from her floor as compared to APs from other floors.}
\label{fig:aps_visible_ex}
\vspace*{-6mm}
\end{figure}
\begin{figure*}[!t]
\includegraphics[width=0.8\linewidth,height=3.4cm]{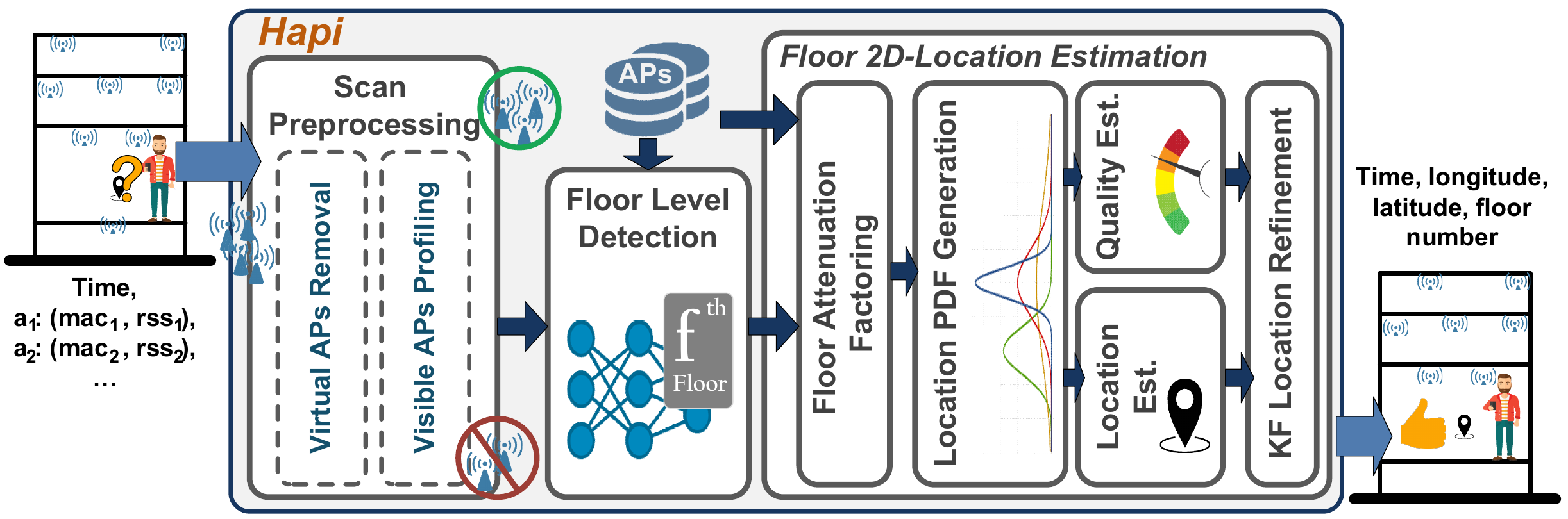}
\vspace*{-2mm}
\caption{The \sys{} system architecture. \sys{} takes only the user's visible APs and their RSS to estimate her pseudo-3D location (the floor number and the longitude and latitude of her 2D-location on that floor).}
\label{fig:arch1}
\vspace*{-3mm}
\end{figure*}

In this paper, we present the \sys{} system as a novel WiFi-based pseudo-3D indoor localization system; i.e. it identifies the user's floor-level and her 2D location on that floor. It is calibration-free, \textbf{only requiring a building's floorplans and its WiFi APs' locations for deployment}. \sys{}, instead of the typical approach of estimating the user location in a 2D setting, works in a pseudo-3D setting incrementally. The key observation here is that while a user gets to hear faraway APs from her floor, she is less likely to hear far APs from other floors as their signals get attenuated by the ceilings/floors as well as the 3D distance between the AP and the user (Figure~\ref{fig:aps_visible_ex}). 
\sys{} takes that into account to improve the user location estimation on her floor. Thus, it starts by identifying the user's floor-level in the building. Next, that floor-level is used, along with the user's visible APs from all floors, in estimating her 2D location. %
In addition, we propose a regression-based algorithm to predict the quality of the estimated location and employ it within a Kalman-Filter (KF) to further refine \sys{}'s location accuracy.

To identify the user floor, we propose \emph{a novel deep-learning based method} that takes the user visible APs as an input and estimates the user's floor. More specifically, we start by limiting the floor search space to the set of floors where the user is most likely located. \emph{This makes our method independent of the building's number of floors}. Then, we extract a set of features based on the APs' pseudo-3D installation location, such as the number of APs per floor and the farthest %
distance between visible APs per floor. Due to WiFi signal propagation characteristics, typically, a user can hear more APs from her floor as compared to other ones and, as highlighted earlier, she can hear more distant APs from her floor as compared to other floors (Figure~\ref{fig:aps_visible_ex}). The extracted WiFi features are then fed into the deep-network for floor prediction. Additionally, to train the network, we present a data balancing and generalization approach that improves the system accuracy and robustness. 

Thereafter, to identify the user 2D-location on her floor, \sys{} considers the user's \emph{visible APs from all floors} and employs her floor-level to compensate for the attenuation the APs' RSS incurs (due to propagation through floors/ceilings). Next, the processed RSS is used to estimate the user location PDF over her entire floor. However, to make the location estimation robust to heterogeneity, \sys{} refrains from using the absolute APs' RSS values while estimating the PDF. Instead, building on the work of \cite{elbakly2016robust,liu2016selective}, we define RSS-Rank that ranks APs' RSS to levels e.g. strong, weak, etc... and use the RSS-Rank in a Gaussian-based method to assess the user proximity probability from each AP across the floor. This helps reduce the effect of the RSS variability, as while the APs' RSS varies (e.g. among devices), their Rank is more robust. For instance, considering a distant AP with a weak RSS, while its absolute RSS may vary, its RSS-Rank is not likely to change from weak to strong.

 We have deployed \sys{} on different android devices (covering a wide-range of models including Samsung, Motorola, LG, OnePlus and Huawei) and conducted extensive experiments \emph{spanning 6 months with 13 subjects in five different multistory buildings with up to 9 floor levels}. Our evaluation results show that \sys{} can identify the user's exact floor up to 95.1\% of the time and 87.5\% of the time overall testbeds. This is better than Locus~\cite{bhargava2015locus} by 52.1\% and 47.7\% respectively. In addition, \sys{} identifies the user's 2D location on that floor with a median accuracy of 3.5m overall testbeds, achieving an improvement of 60.7\% and 76.0\% over IncVoronoi~\cite{elbakly2016robust} and Locus~\cite{bhargava2015locus} respectively. We believe %
 \sys{}'s performance, over such large scale testbeds,
  marks a significant milestone towards achieving a true ubiquitous indoor localization system.

In summary, our main contributions are summarized as follows:
\begin{itemize}
\item We propose \sys{}---a novel calibration-free WiFi-based indoor localization system that takes into account visible APs from all floors to estimate the user floor-level and her 2D location on that floor.
\item We present a deep-learning-based floor-level estimation method which only requires the building's WiFi APs location for training and deployment. \emph{We believe this method can work independently %
for WiFi-based floor determination}.
\item We present the details of an RSS-Rank probabilistic-based algorithm that estimates the user location using her current floor-level and visible APs from all floors.
\item We present a regression-based method for predicting \sys{}'s user-location estimates' quality. We employ it within a KF to further refine the system's accuracy.
\item We implement the system on a wide-range of Android devices and thoroughly evaluate it in 5 different multi-story buildings with 13 subjects over 6 months. In addition, we give a side-by-side comparison with two state-of-the-art calibration-free systems.
\end{itemize}

The rest of the paper is organized as follows: Section~\ref{sec:overview} presents an overview of the \sys{} system architecture. Then, we give the details of its floor-level and  2D location estimation in Section~\ref{sec:floorLoc} and~\ref{sec:locEst} respectively. Section~\ref{sec:eval} describes our implementation and evaluation of \sys{}. Section~\ref{sec:relatedWork} discusses related work. Finally, we give concluding remarks with directions for future work in Section~\ref{sec:conc}.

\section{The \system{} Architecture}\label{sec:overview}
\begin{figure*}[!t]
\includegraphics[width=0.8\linewidth,height=3.4cm]{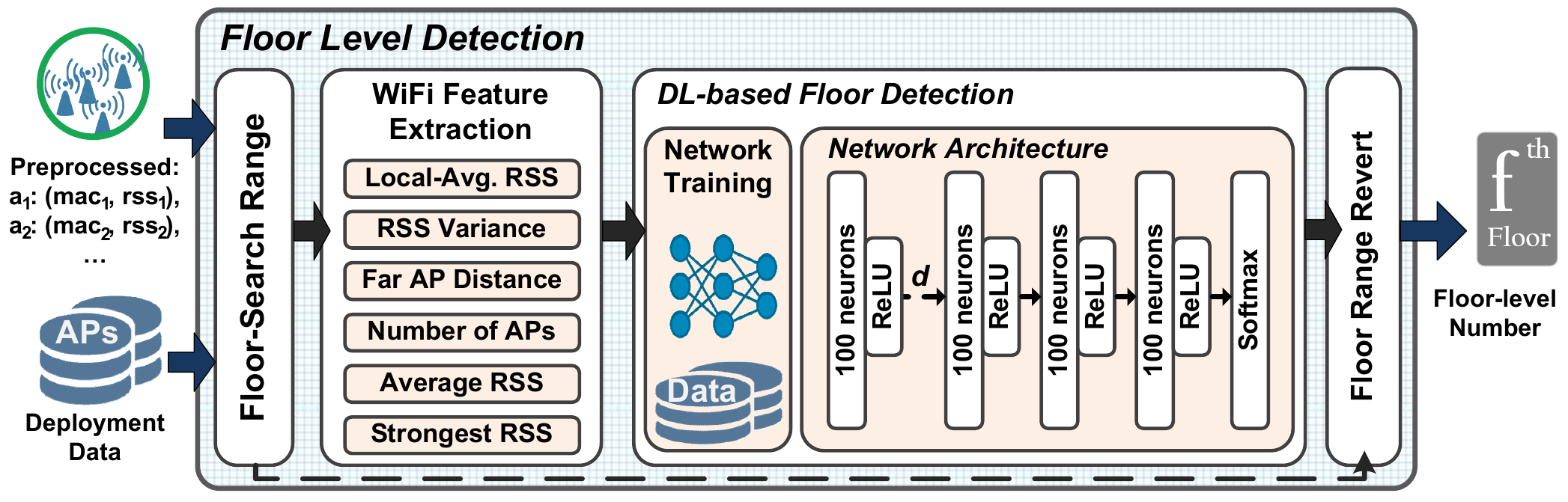}
\vspace*{-2mm}
\caption{Overview of \sys{}'s \emph{Floor Level Detection} module.}
\label{fig:floor_det_ov}
\vspace*{-3mm}
\end{figure*}

Figure~\ref{fig:arch1} shows the \sys{} system architecture. To deploy \sys{} in a building, only a list of the building's installed WiFi APs' MAC addresses and installation locations is fed to the system along with the building's floor plans. To localize a user, the system takes %
a time-stamped WiFi scan list which consists of the user's visible APs' MAC addresses and their received signal strengths (RSSs). \sys{} has three main modules: the \emph{Scan Preprocessing} module, the \emph{Floor Level Detection} module and the \emph{Floor 2D-Location Estimation} module. %

\subsection{Scan Preprocessing Module}
For a given WiFi scan list, we start by mapping any virtual APs to their physical AP MAC address. A virtual AP is a multiplexed installation of a single physical AP so that it presents itself as multiple discrete APs to the wireless LAN clients. The mapping function should be provided at deployment with the building's WiFi network information. For example, in our testbed buildings, the last hexadecimal digit for a physical AP's MAC address is 0 and has different values for its virtual APs. Thus, we mask the last hexadecimal digit to 0 to map all virtual APs to their physical AP's MAC address. Then, we set the physical AP's RSS to the average RSS of its virtual APs. %

In addition, \sys{} creates a profile for the user's surrounding APs and their RSS through using a sliding window of the WiFi scans. This is intuitive as within a short period, in typical indoor scenarios, the user is less likely to move far from her current location and the window can show more APs in the user's vicinity---a single WiFi scan may only show a subset of the nearby APs. %
  Thus, for a window of size $N$, a WiFi profile ($p_t^N$) of the user's surrounding WiFi network at time $t$ is constructed by taking the union of all visible APs in $N$ scans. If an AP is visible in multiple scans, its strongest RSS is used in $p_t^N$. More formally, $p_t^N$ is defined as follows:
\vspace*{-3mm}
\begin{equation}
\small
\vspace*{-3mm}
\begin{split}
p_t^N & = \bigcup_{i\; = \;\textit{max}(0,t-N)}^{t} s_i \\
& = \{ a_m: a_m = (\textit{mac}_m, \textit{rss}_m),  \exists s_i \supset a_m , i \in [\textit{max}(0,t-N):t] \}
\end{split}
\end{equation}
Where $s_i = \{a_1,...,a_m,...\}$ is the WiFi scan list at time $i$ with an AP $a_m = (\textit{mac}_m, \textit{rss}_m)$ is defined by its MAC ($\textit{mac}_m$) and RSS ($\textit{rss}_m$).

\subsection{Floor Level Detection Module}
This module is responsible for detecting the user's floor level. It takes as an input, a visible APs profile $p_t^{N_f}$.  We show the $N_f$ size effect in Section~\ref{sec:eval}. The module extracts a set of features, from $p_t^{N_f}$, that highlight the WiFi characteristics at each floor using the APs' pseudo-3D locations. For example, it extracts the number of APs from every floor as the user is more likely have more visible APs from her floor as opposed to other floors due to signal attenuation (e.g. Figure~\ref{fig:aps_visible_ex}). Then, these features are used to estimate the user floor $f_t$ through a deep neural network. Furthermore, the module incorporates RSS variability for data balancing and generalization while training the network to achieve high robust accuracy. The details of operation of this module are discussed in Section~\ref{sec:floorLoc}.

\subsection{Floor 2D-Location Estimation Module}
This module is responsible for estimating the user 2D location within her floor level. It takes as an input, a visible APs profile $p_t^{N_l}$ and the user's estimated floor level $f_t$.  We show the $N_l$ size effect in Section~\ref{sec:eval}. \sys{} estimates the user location using a novel probabilistic method as we discuss in detail in Section~\ref{sec:locEst}.  

Note that the \emph{Floor 2D-Location Estimation} module has $N=N_l$ which is different from the $N_f$ used in the \emph{Floor Level Detection} module. This is intuitive as the user will typically be moving/walking for a while in the same floor as compared to her location within that floor. Thus, we expect $N_f$ to be longer than $N_l$. %

\section{Floor-Level Detection}\label{sec:floorLoc}
A key part of \sys{}'s localization algorithm is identifying the user's current floor. The \textit{Floor-Level Detection} module takes $p_t^{N_f}$ and outputs the user's floor estimate $f_t$. The module starts by finding the floors' range where the user is most-likely located out of the entire building. Then, WiFi features are extracted and are fed to a Deep Neural Network (DNN) to identify the user's floor. Figure~\ref{fig:floor_det_ov} shows an overview of the \textit{Floor-Level Detection} module.

\subsection{Floor-Search Range}

\begin{figure*}[!t]
\vspace*{-1mm}
\begin{minipage}{0.31\linewidth}
\includegraphics[width=0.95\linewidth]{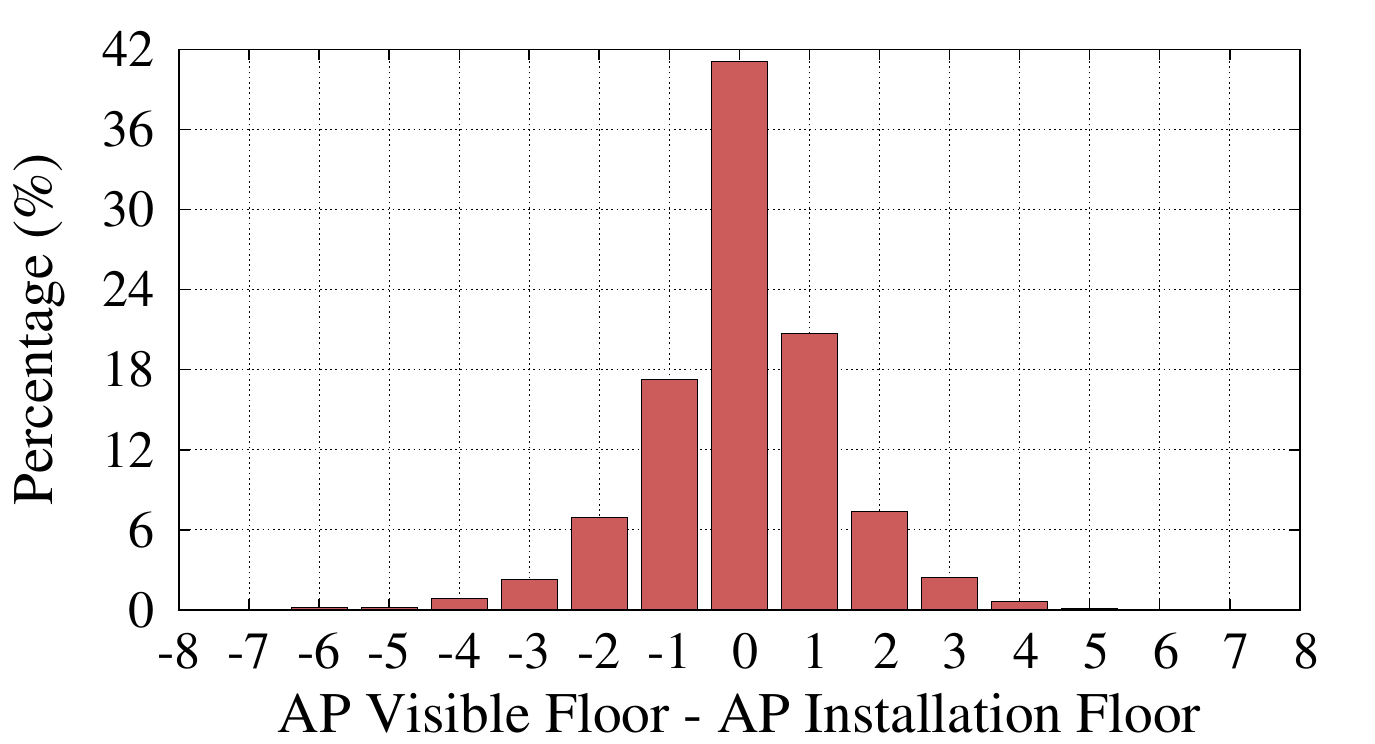}
\caption{Histogram of the difference between the floor where an AP is visible and its installation floor for APs in our experimental data.} %
\label{fig:aps_range}
\end{minipage}
\hspace{1mm}
\begin{minipage}{0.35\linewidth}
\center
\subfigure[\underline{User location:} \textbf{1st floor}]{
 \includegraphics[width=0.46\linewidth]{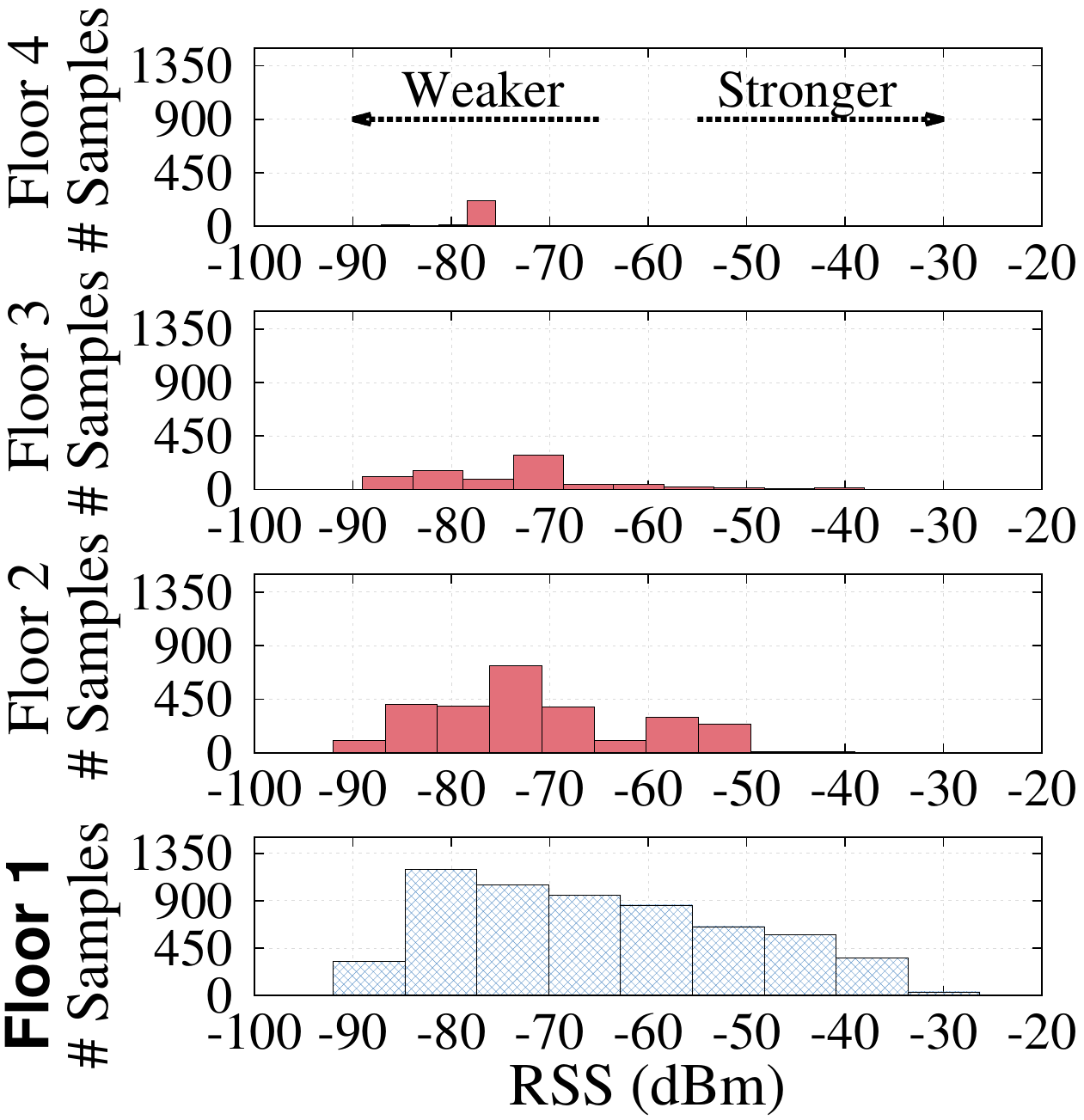}
      \label{fig:wifi_ex1}
}
\subfigure[\underline{User location:} \textbf{3rd floor}]{
      \includegraphics[width=0.46\linewidth]{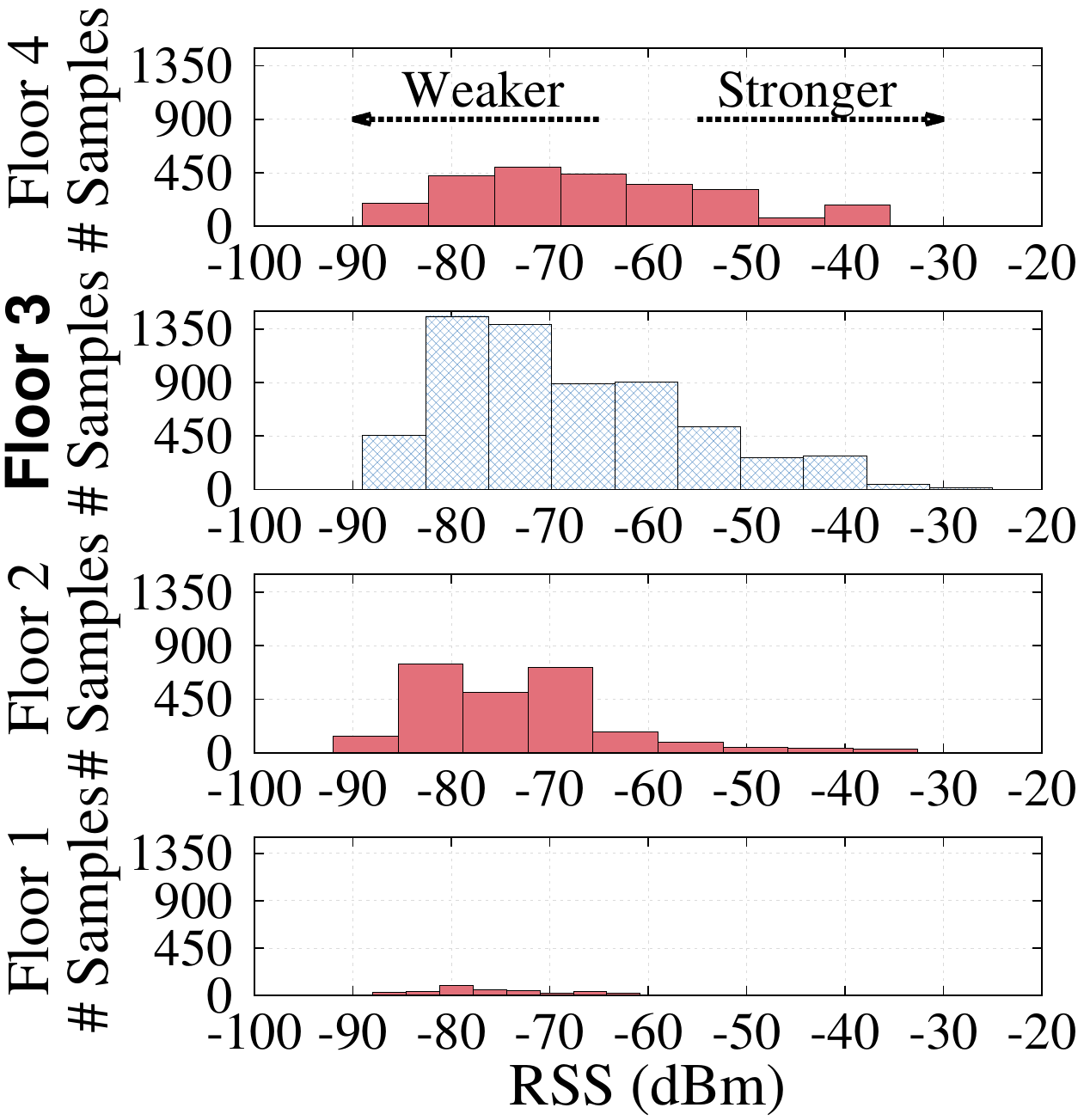}
      \label{fig:wifi_ex2}
    }
\vspace*{-4mm}
\caption{Histogram of visible APs' RSS for users walking in the 1st floor vs 3rd floor of the testbed Bldg. 115.%
}
\label{fig:floors_hist}
\end{minipage}
\hspace{0.5mm}
\begin{minipage}{0.29\linewidth}
	\centering	
\includegraphics[width=0.95\linewidth]{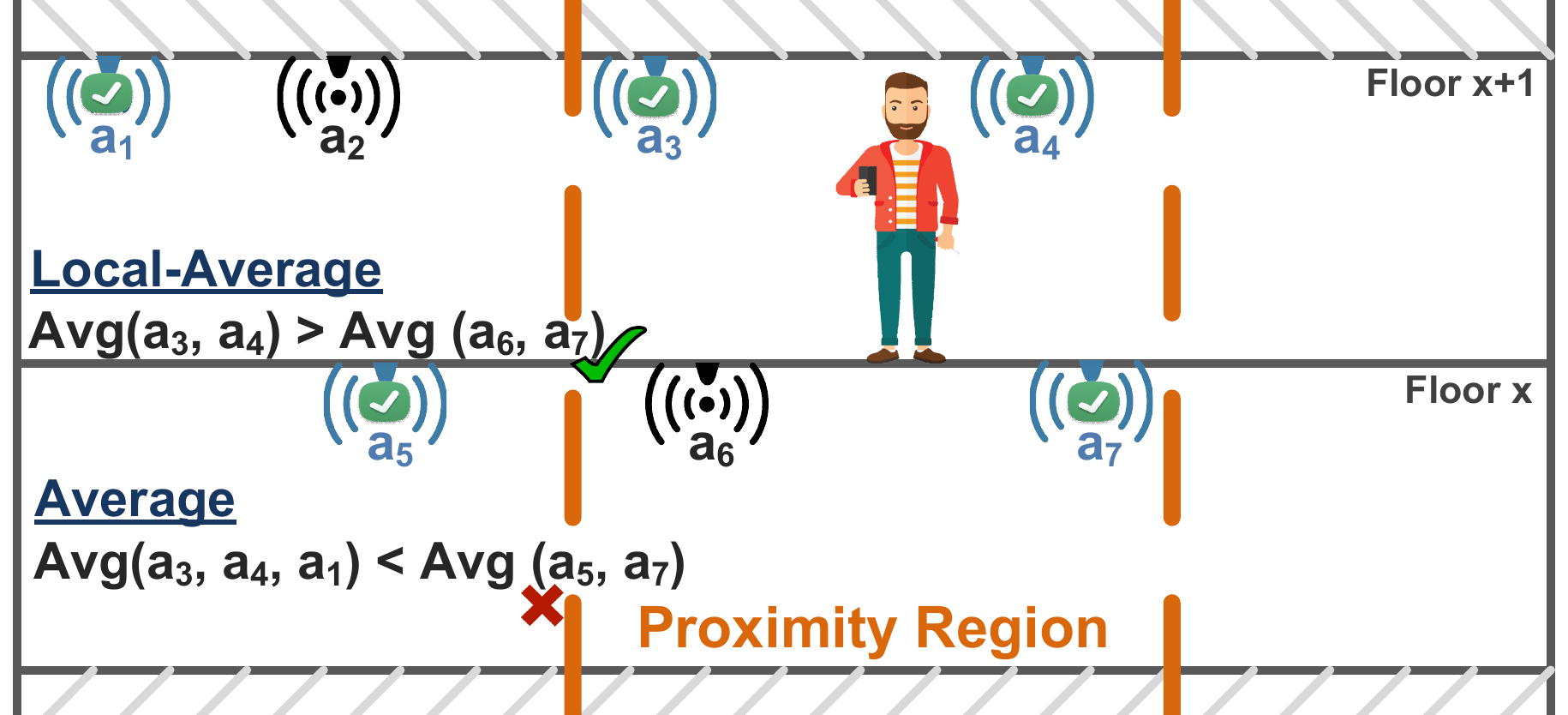}
\vspace*{-3mm}
\caption{Example illustrating Local-Average Signal Strength and its proximity region vs. Average Signal Strength computation. Visible APs are marked with a tick.%
}
\label{fig:loc_avg_illu1}
\end{minipage}
\vspace*{-4mm}
\end{figure*}

While buildings can consist of any number of floors, WiFi APs have limited reception range.  For example, Figure~\ref{fig:aps_range} shows a histogram of the difference between the floor where an AP is visible and its installation floor for all APs in our experiments data which includes up to 9 floor buildings (Section~\ref{sec:eval}). We can see that typically, 98\% of the time, an AP can only be heard within four floors from its installation floor. We leverage this to limit the search range of the user-floor finding algorithm. Specifically, for a user with WiFi profile $p_t^N$ in a building with $\mathcal{F}$ floors, \sys{} limits its floor-search range to a subset $F \in \mathcal{F}$ of up to $w$ consecutive floors. $w$ is a system parameter that defines maximum number of floors to search for the user in. Thus, $|F| \leq |\mathcal{F}|$ and $|F| \leq w$. This enables our floor-finding algorithm to work on $w$ floors, independently from the building's number of floors $|\mathcal{F}|$, \textbf{allowing \sys{} to be deployable in any building with any number of floors}. Due to signal propagation characteristics, APs closer to the user will have stronger RSS. Thus, we define $F$ as the consecutive subset of floors where the user is covered with the strongest overall RSS. The search range $F$ start index $r_1$ is computed as follows:
\vspace*{-1mm}
\begin{equation}
\vspace*{-1mm}
r_1 = \argmax_{f \in [1:|\mathcal{F}-w|]} \sum_{i = f}^{f+w} \sum_{\forall a_m \in A_i} \textit{rss}_m
\label{eq:range}
\end{equation}
Where $A_i$ %
is the set of APs installed in floor $i$. 
Note that, APs from floors outside $F$ (if any) is removed as they represent outlier weak APs. We set $w$ to 4 based on our analysis (Figure~\ref{fig:aps_range}) and evaluation.

\subsection{WiFi Feature Extraction}
For a WiFi profile $p_t^N$, we extract a set of features from each of the $w$ floors. %
First, we extract features based on the building's APs' installation floor: The number of access-points, the strongest signal strength, the average signal-strength and the signal strength variance. In addition, we propose features that are based on the APs' pseudo-3D installation location: The local-average signal strength and the farthest access-point distance per floor. We describe each in detail throughout this section.

\subsubsection{Number of Access-points}
WiFi accesspoints have a limited range (around $\sim$100m) and this range decreases with attenuation from walls and other indoor clutter. Thus, we expect to have more visible APs in the user current floor and the number should decrease as we go further away (Figure~\ref{fig:aps_visible_ex}). For instance, Figure~\ref{fig:floors_hist} shows the histogram of visible APs' RSS for users walking in testbed Bldg. 115 1st and 3rd floors (Section~\ref{sec:eval}). We can see that more APs were visible in the users' floors (1st and 3rd respectively). 

Therefore, our first extracted feature is the number of visible APs per floor ($\textit{Num}_f$) in the WiFi profile $p_t^N$. For each floor $f \in F$:%
\vspace*{-1mm}
\begin{equation}
\textit{Num}_f = |A_f|
\end{equation}
\subsubsection{Strongest Signal Strength}
WiFi signals get attenuated, by distance and obstacles (e.g. floors/ceilings), leading to weaker RSS. %
Thus, we expect APs in the user's floor to have a relatively strong RSS compared to APs in distant floors. Going back to the visible APs' RSS histogram example in Figure~\ref{fig:floors_hist}, we can see that there is a relatively high number of APs with strong RSS ($ > -60 $dBm) in the user's floor (1st and 3rd respectively) and this number decreases as we go up/down. Thus, our second extracted feature is the strongest RSS per floor ($\textit{Str}_f$) in the WiFi profile $p_t^N$. For each floor $f \in F$:%
\vspace*{-1.5mm}
\begin{equation}
\vspace*{-1mm}
\textit{Str}_f = \max_{a_m \in A_f} \textit{rss}_m %
\end{equation}
\subsubsection{Average Signal Strength}\label{sec:avg}
Other than decreasing the maximum RSS, as all signals coming from APs in distant floors get attenuated, their average RSS decreases as well. We can see in Figure~\ref{fig:floors_hist} that the user's floor has higher ratio of APs with strong RSS ($ > -60 $dBm) thus it can have higher visible APs' RSS on average. 
Therefore, the next feature we extract is the average visible APs' RSS per floor ($\textit{Avg}_f$) in the WiFi profile $p_t^N$. For each floor $f \in F$:%
\vspace*{-2mm}
\begin{equation}
\vspace*{-1mm}
\textit{Avg}_f = \frac{1}{|A_f|}\sum_{a_m \in A_f} \textit{rss}_m %
\end{equation}
\subsubsection{Signal Strength Variance}
While all signals from APs' installed at distant floors get attenuated by the ceilings/floors between the APs and the user, this is not the case for APs in the user's floor. APs installed at the user's floor, based on their proximity to the user, can have very strong or weak RSS.
 For instance, in Figure~\ref{fig:floors_hist}, whereas APs at distant floors exhibit a predominantly weak RSS, the user-floor's APs have both weak and strong RSS. Leading to a higher RSS variance in the user's floor as compared to distant ones. %
 Hence, we extract the RSS variance per floor ($\textit{Var}_f$) in the WiFi profile $p_t^N$. For each floor $f \in F$:%
\vspace*{-2mm}
\begin{equation}
\textit{Var}_f = \frac{1}{|A_f|}\sum_{a_m \in A_f} (\textit{rss}_m - \textit{Avg}_f)^2 %
\end{equation}

\subsubsection{Local-Average Signal Strength}\label{sec:locavg}

\begin{figure*}[!t]
\begin{minipage}{0.34\linewidth}
\center
\includegraphics[width=\linewidth,height=3.1cm]{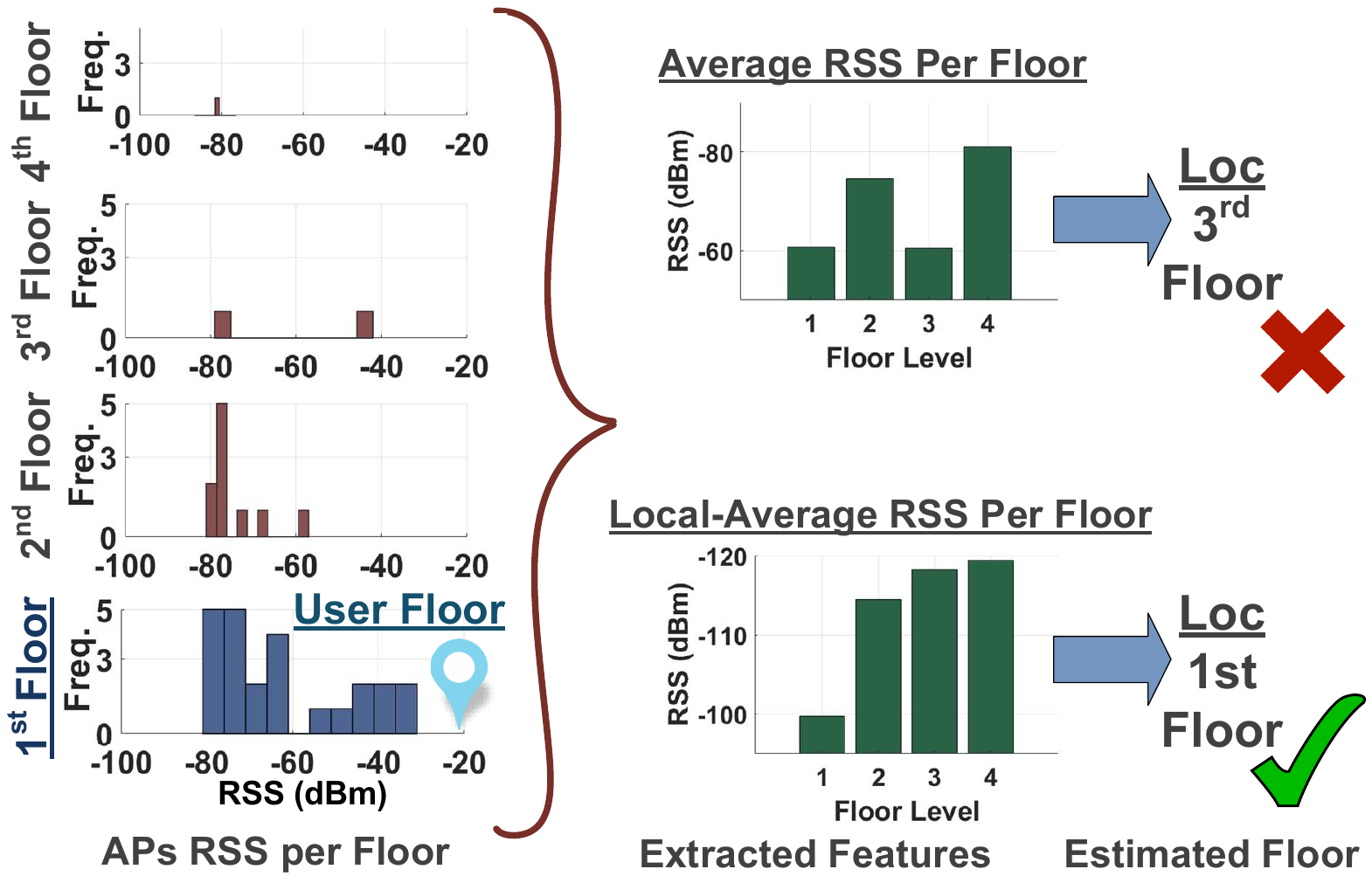}
\vspace*{-4mm}
\caption{Example comparing the \emph{Average Signal Strength} to the \emph{Local-Average Signal Strength} features on user-floor detection.
}
\label{fig:loc_avg_ex1}
\end{minipage}
\hspace*{1.5mm}
\begin{minipage}{0.64\linewidth}
\center
\subfigure[\textbf{Without} Floor Attenuation Factoring]{
\includegraphics[width=0.44\linewidth,height=2.7cm]{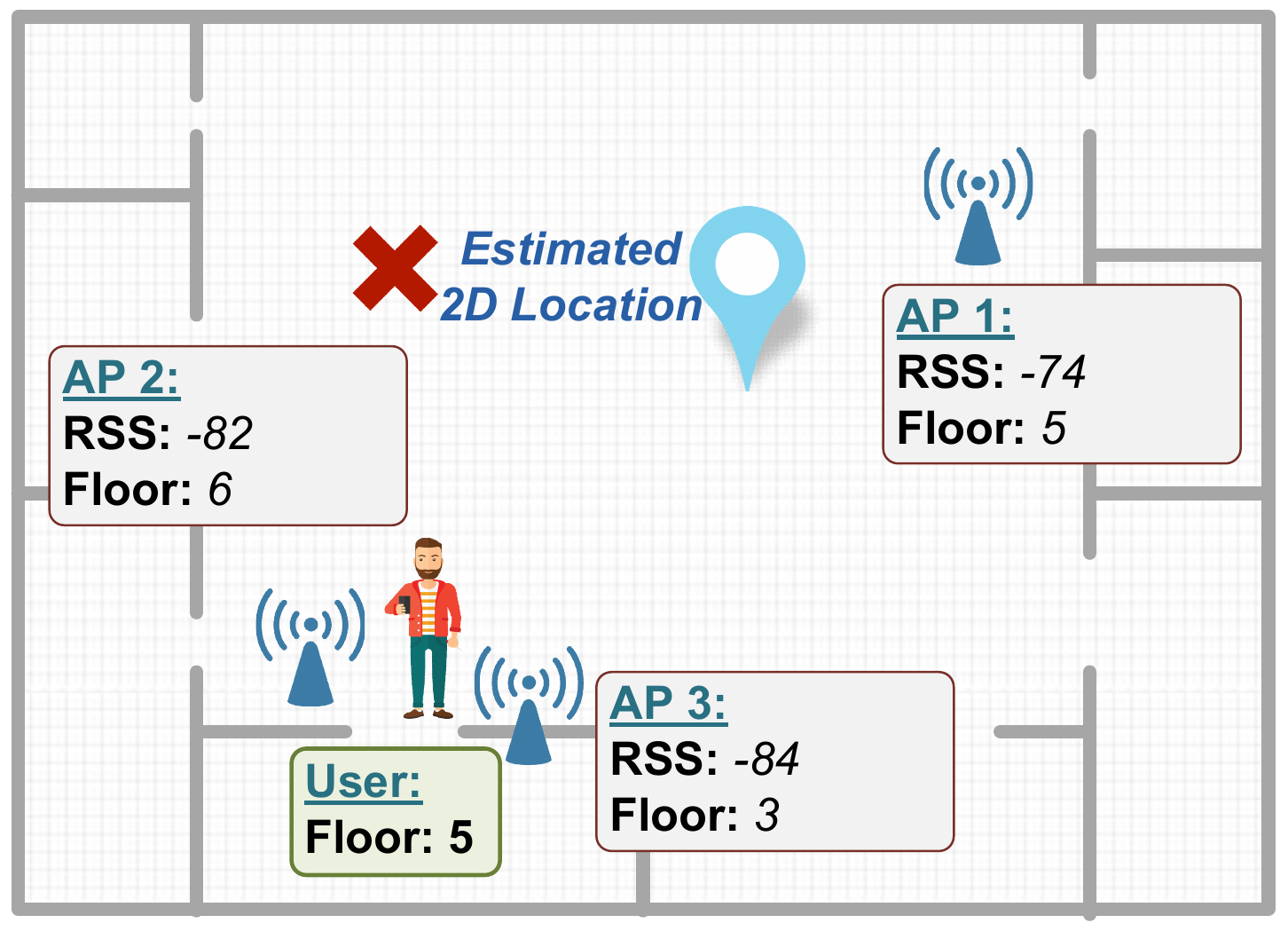}
\label{fig:faf_ex1}
}
\subfigure[\textbf{With} Floor Attenuation Factoring]{
 \includegraphics[width=0.44\linewidth,height=2.7cm]{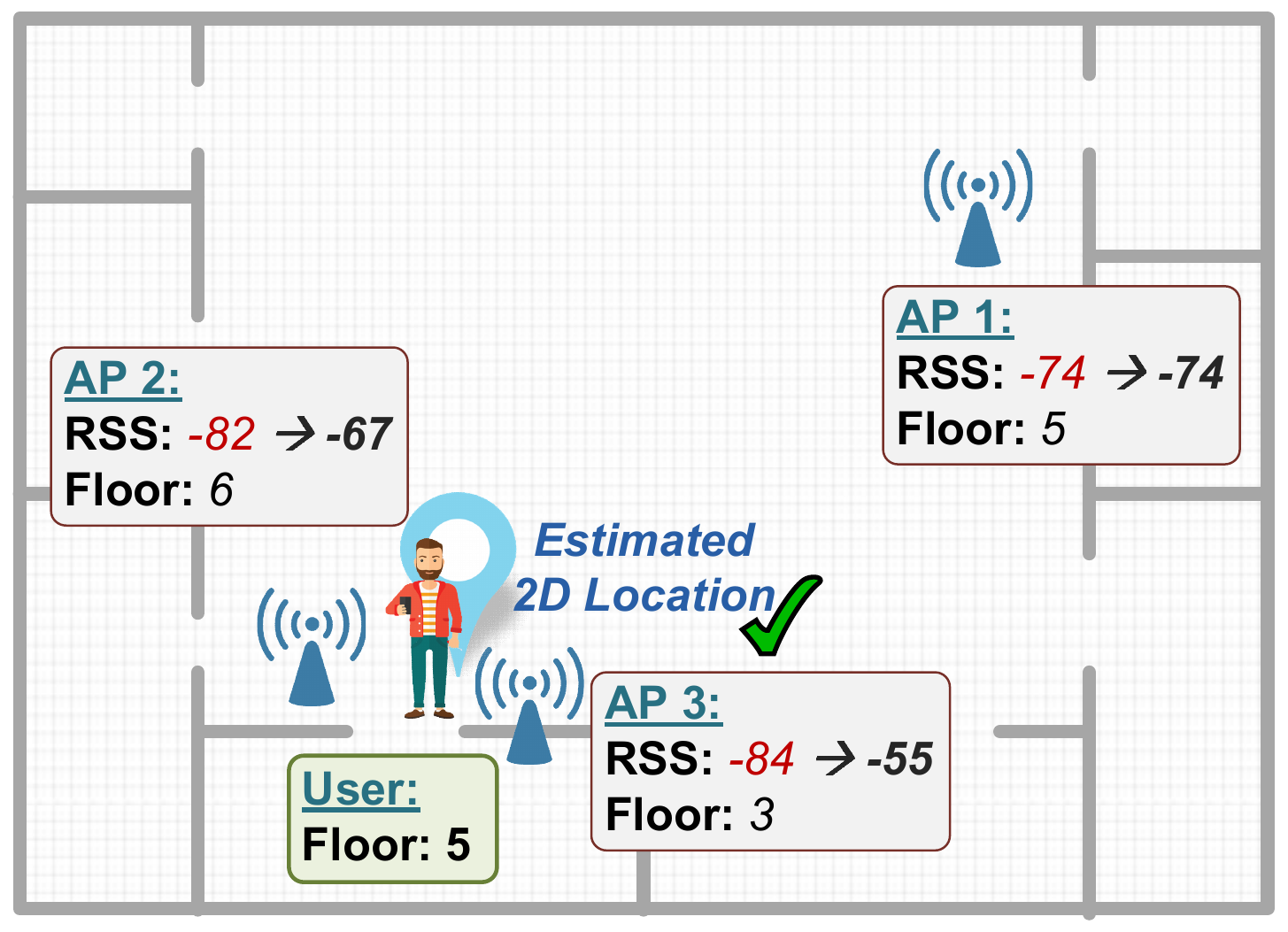}
\label{fig:faf_ex2}
}
\vspace*{-4mm}
\caption{Example showing the effect of the APs' installation floor on their RSS and \sys{}'s 2D location estimation with and without the \emph{Floor Attenuation Factoring} module. The user is located in floor 5.}
\label{fig:faf_ex}
\end{minipage}
\vspace*{-3mm}
\end{figure*}

One problem with taking the average of all visible APs' signal strength is that, the user's actual floor can have high RSS variance (as discussed in previous features), making the average not representative. To overcome this problem, we propose the \emph{local-average signal strength} feature which utilizes the installed APs' pseudo-3D location. Particularly, rather than computing the visible APs' average RSS within the entire floor, we compute a local-average RSS for all installed APs within a \emph{proximity region} (Figure~\ref{fig:loc_avg_illu1}). If an AP in the region is not visible, we assume a weak RSS of $-100dBm$. Figure~\ref{fig:loc_avg_ex1} compares the extracted \emph{Average Signal Strength} and \emph{Local-Average Signal Strength} features and shows how using the local-average can reduce the user's floor high RSS variance problem through taking the WiFi network deployment into account. The proximity region $\mathcal{R}_\delta$ is defined by triangulating the WiFi profile $p_t^N$ APs' 2D locations and drawing a circle with radius $\delta$ around it. Then, for each floor $f \in F$, the local-average RSS ($\textit{LocAvg}_f^\delta$) is computed as follows: %
\vspace*{-1mm}
\begin{equation}
\vspace*{-2mm}
\textit{LocAvg}_f^\delta = \frac{1}{|I_f^\delta|}\sum_{a_m \in I_f^\delta} \textit{rss}_m %
\end{equation}
\vspace*{-1mm}
Where %
$I_f^\delta$ is the set of APs at floor $f$ in $\mathcal{R}_\delta$. %
 We set $\delta$ to 30m, 80m and the farthest distance between visible APs ($\alpha$) to cover different regions granularity.  We analyze the effect of $\delta$ in Section~\ref{sec:eval}.
 \begin{equation}
 \alpha = \max_{a_m \in p_t^N, a_l \in p_t^N} \textit{dist}(a_m,a_l)
 \end{equation}
 Where $\textit{dist}(a_m,a_l)$ is the Haversine distance between $a_m$ and $a_l$. %
\subsubsection{Farthest Access-point Distance}

As discussed earlier, the user is more likely to hear distant APs from her floor as compared to other floors (e.g. Figure~\ref{fig:aps_visible_ex}). Thus, to capture this observation, we extract the farthest pairwise visible APs distance ($\textit{Far}$) at each floor as our last feature. %
For each floor $f \in F$: %
\begin{equation}
\textit{Far}_f = \max_{a_m  \in A_f,a_l \in A_f} \textit{dist}(a_m,a_l)
\end{equation}

\subsection{Deep Learning-based Floor Detection}
We formulate the floor detection problem as finding which of the $w$ search-range floors is the user's actual floor (${f_t}'$) and propose a Deep Neural Network (DNN) to solve it. The extracted WiFi features from each of the $w$ floors are fed into the DNN and it outputs ${f_t}'$. %
 Our DNN has four fully connected layers each with 100 neurons. We employ two design rules in the DNN architecture: (1) we have used a dropout layer with a dropout ratio of 0.25 (25\% chance of setting a neuron's output value to zero) after the first fully connected layer~\cite{srivastava2014dropout} to improve the network generalization capability; and (2) the non-linearity for the four fully connected layers is fulfilled by Rectified Linear Unit (ReLU)~\cite{nair2010rectified} as the activation function to prevent saturation of the gradient when the network is deep. The network ends with a softmax layer to predict the floor-level probability of each of the search-range $w$ floors and ${f_t}'$ is set as the most probable one. A basic layer block is formalized as:
\begin{equation}
y =  \textbf{W} \cdotp \textbf{x} + \textbf{b}
\end{equation}
\begin{equation}
h = \textit{ReLU}(y)
\end{equation}

Where $\textbf{W}$ denotes the weights vector, $\textbf{x}$ is the features vector, $\textbf{b}$ is the bias. In the layer where we add the dropout (with dropout ratio $d$), $\textbf{x}$ is substituted by $\tilde{\textbf{x}}$ where:
\begin{equation}
\tilde{\textbf{x}} =  f_{\textit{dropout},d}(\textbf{x})
\end{equation}
In the first layer, $\mathbf{x}$ is the extracted WiFi features vector. Figure~\ref{fig:floor_det_ov} shows the neural network architecture. 

\subsubsection{Training: Data Generalization and Augmentation}
To deploy \sys{}'s Floor-level Detection module in a building, \textbf{no WiFi data collection is required from that building}. Instead, to train the network, we emulate the deployment-building's WiFi network%
. In addition, we generalize and balance the training data's WiFi scans to prevent overfitting. Specifically, for each training data instance, we start by sampling the APs per floor to match the number of Aps per floor in any randomly selected consecutive $w$ floors (search-range) of the deployment-building. Then, we perturb the sampled-APs' RSS by adding a Gaussian random amount with zero mean and standard-deviation $\sigma$ proportional to the degree of noiseness we want to add to the model. Based on our analysis of the APs' RSS variability, we set $\sigma$ to $1.5$. Finally, to balance the data, we repeat the Gaussina-random perturbation to equalize the number of instances in each floor. 
We evaluate the effect of our training methodology in Section~\ref{sec:eval}.

\subsubsection{Training: Optimization}
We use a categorical cross entropy loss function and the Adam optimization algorithm ($\beta_1 \; = \; 0.9$, $\beta_2 \; = \; 0.999$, $\epsilon \; = \; 10^{-8}$, a batch size of 10 and a learning rate of $10^{-3}$)~\cite{kingma2014adam} for training the network.

\subsection{Floor-Range Revert}
The DNN estimates the user's most probable floor ${f_t}'$ out of the search-range $w$ floors. Then, we revert ${f_t}'$ back to its actual floor-level number $f_t$ out of the building's $\mathcal{F}$ floors using the search-range 1st index $r_1$ (Equation~\ref{eq:range}): $f_t = {f_t}' + r_1$.
\begin{figure*}[!t]
\begin{minipage}{0.55\linewidth}
\vspace*{-1mm}
\center
\subfigure[{\underline{Rank:} Very Strong}]{
 \includegraphics[width=0.31\linewidth]{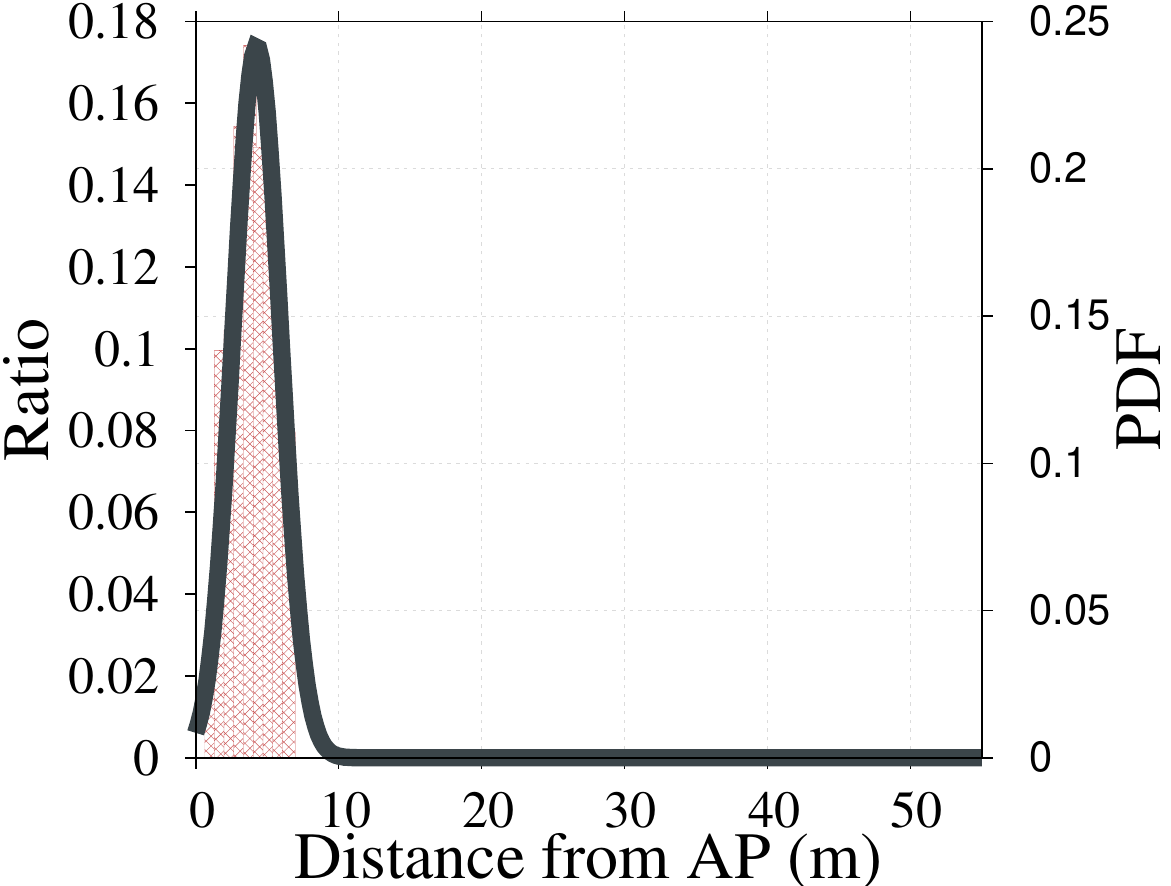}
      \label{fig:n_ex_40}
}
\hspace*{-2mm}
\subfigure[{\underline{Rank:} Moderate}]{
 \includegraphics[width=0.31\linewidth]{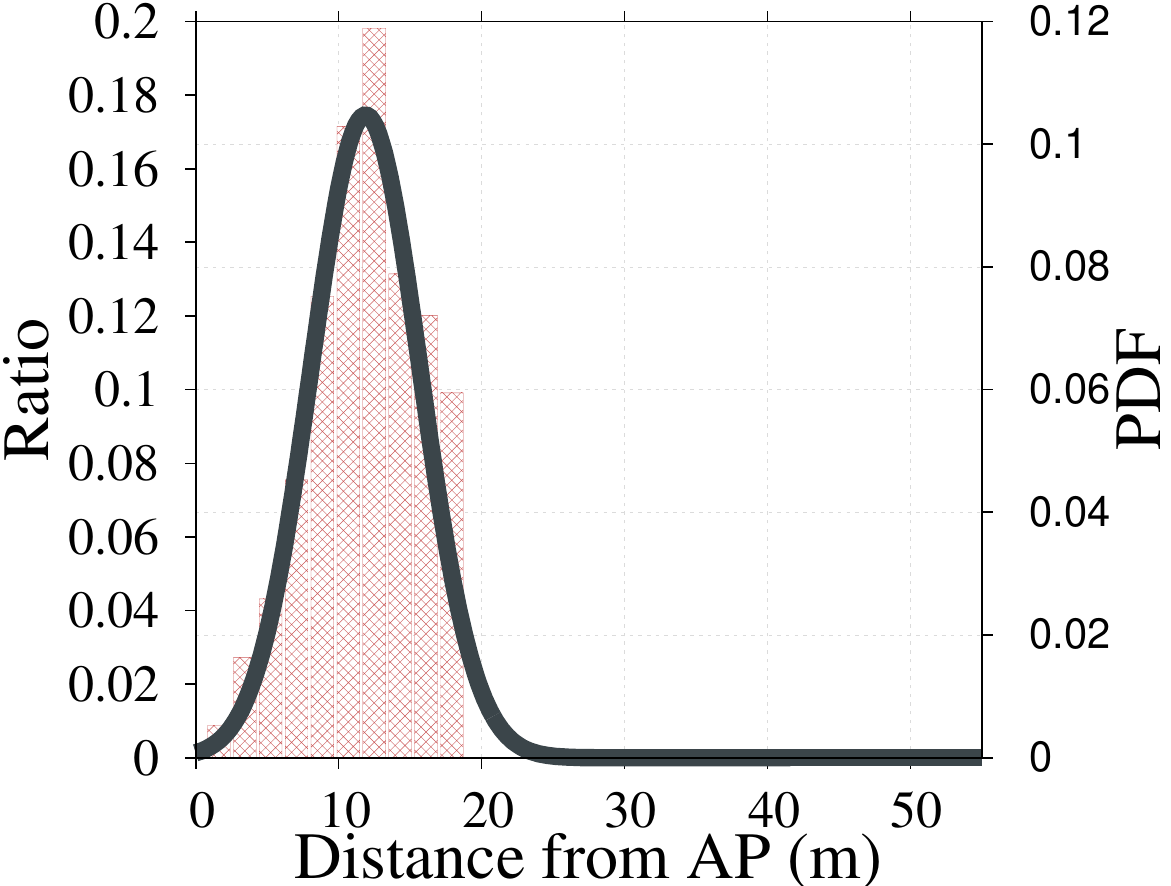}
      \label{fig:n_ex_60}
}
\hspace*{-2mm}
\subfigure[{\underline{Rank:} Weak}]{
 \includegraphics[width=0.31\linewidth]{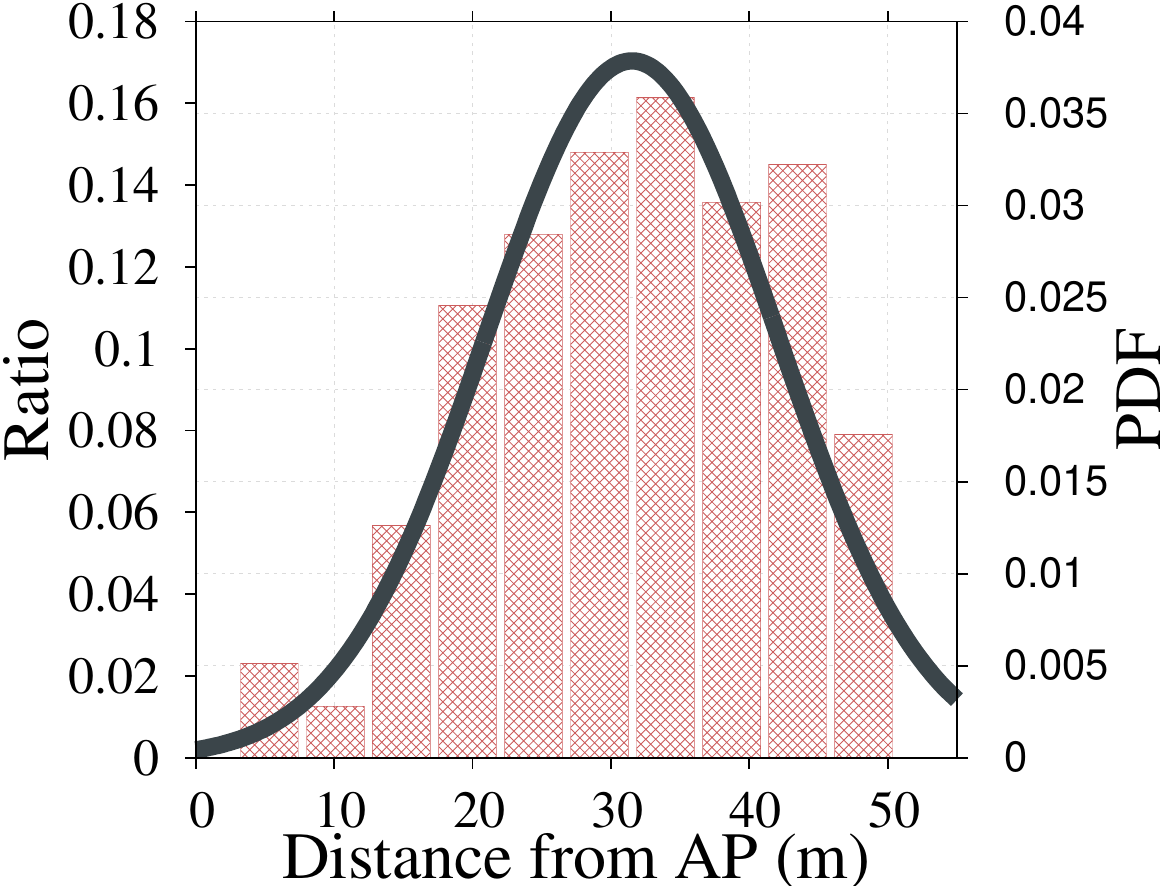}
      \label{fig:n_ex_80}
}
\vspace*{-4mm}
\caption{Histogram of user's distance from visible APs with different RSS-Ranks (very strong, moderate, weak) collected in testbed Bldg. 115 and \sys{}'s RSS-Ranks' Gaussian models' PDF.}
\label{fig:n_ex}
\end{minipage}
\hspace*{1mm}
\begin{minipage}{0.43\linewidth}
\center
\includegraphics[width=0.95\linewidth]{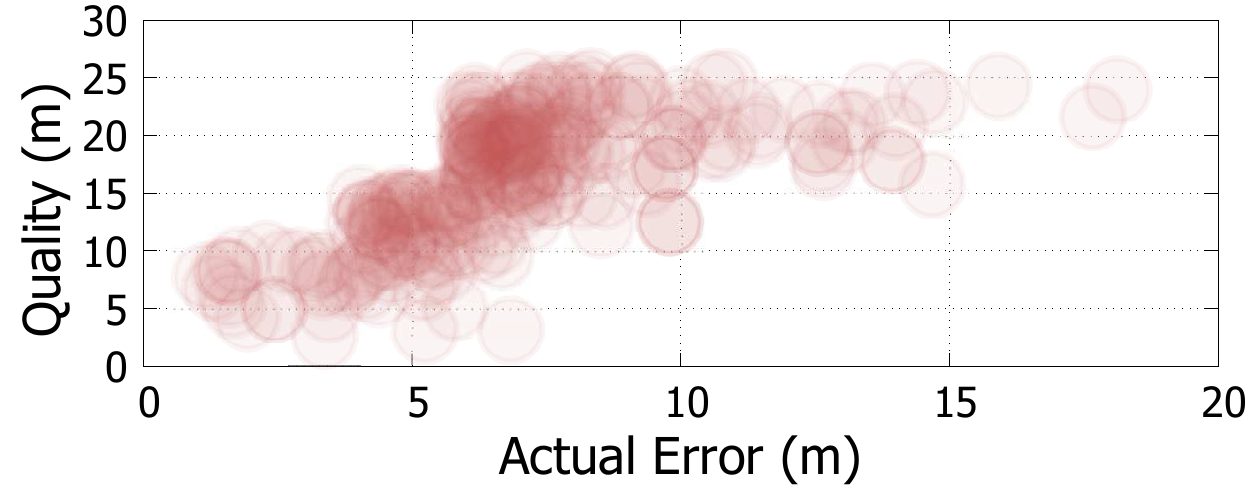}
\vspace*{-4mm}
\caption{Example comparing the estimated-location actual-error and its predicted quality.}
\label{fig:zeft}
\end{minipage}
\vspace*{-4mm}
\end{figure*}
\section{Floor 2D-Location Estimation}\label{sec:locEst}
In this section, we provide \sys{}'s \emph{Floor 2D Location Estimation} module details. It takes $p_t^{N_l}$ along with the user's detected floor $f_t$, and outputs her 2D location on $f_t$%
. The module starts by preprpocessing the APs to compensate for the attenuation their RSS incur due to their installation floor. Then, we employ a novel RSS-Rank Gaussian model to estimate the user location PDF across her entire floor and assess the user's most probable location. Finally, a regression-based model is proposed to predict the quality of the user-location and it is employed within a KF to further refine that location.
 
\subsection{Floor Attenuation Factoring}\label{sec:faf}
While users typically have visible APs from a number of floors nearby her actual floor, those APs' RSS get attenuated by the floors/ceilings in between.
 Although this attenuation helps in identifying the user floor-level (as shown in Section~\ref{sec:floorLoc}), it leads to confusion when estimating the user 2D location on her floor (Figure~\ref{fig:faf_ex1}). To address this issue, we build on the well-known Wall Attenuation Factoring method, which have been used in indoor localization systems e.g.~\cite{bahl2000radar,elbakly2016robust}, and compensate for the APs' installation floor through the Floor Attenuation Factoring~\cite{seidel1992914} (Figure~\ref{fig:faf_ex2}). \sys{} leverages the user floor estimate $f_t$ and the AP installation floor to estimate the number of ceilings/floors between the user floor and the AP. In
particular, for a user at floor $f_t$ with a WiFi visible APs profile $p_t^N$, every AP $a_m \in p_t^N$ is updated as follows:
\begin{equation}
\vspace*{-1.5mm}
\textit{rss}_m = \textit{rss}_m + W_f \times |(f_t - f_m)|
\end{equation}
Where $f_m$ is the AP $a_m$ installation floor and $W_f$ is a constant parameter representing the signal attenuation due to floors/ceilings. We evaluate the effect of $W_f$ on performance in Section~\ref{sec:eval}.

\subsection{Location PDF Generation}

Next, using the processed WiFi profile $p_t^N$, \sys{} estimates the PDF of the user location (denoted by $\mathcal{L}(p_t^N, \ell_{(x,y)})$) over the entire floor using a novel RSS-Rank Gaussian-based method. $\ell_{(x,y)}$ is the longitude and latitude location variable. A key observation noted in related work is that using the exact RSS value to estimate the user-location leads to device and environment heterogeneity problems and can limit the ubiquity of the proposed model~\cite{elbakly2016robust}. To address this issue, related work systems relied on the relative order between the APs such as IncVoronoi~\cite{elbakly2016robust} and Liu et al.~\cite{liu2016selective}. However, while this approach helps with the heterogeneity and model overfitting, it has two main issues: First, when APs are having close RSS values (e.g. -76 and -77), ordering them or saying that you are closer to any of them is dubious. Second, when using the relative order, you are losing  part of the information in $p_t^N$ (i.e. the APs' signal strength level). Thus, \sys{} extends on related work and normalizes the RSS values to RSS-Ranks. RSS-Ranks rates APs based on their RSS to different ranks such as strong, moderate and weak. Using ranks gives the benefits of the APs ordering (we are not using the RSS absolute value). Yet, we are still relatively using its strength level through the normalization ranks. 

To estimate the user-location PDF $\mathcal{L}(p_t^N, \ell_{(x,y)})$ (Equation~\ref{eq:pdf}), we model the probability of being at distance $d$ from an AP with RSS-Rank $z$ using a Gaussian distribution $\mathcal{N}(\mu_z,\sigma_z)$.
\begin{equation}
\mathcal{L}(p_t^N, \ell_{(x,y)}) = \sum_{a_m \in p_t^N} \mathcal{N}(\textit{dist}(a_m,\ell_{(x,y)}),\mu_z,\sigma_z) %
\label{eq:pdf}
\end{equation}
Where $\mathcal{L}(p_t^N, \ell_{(x,y)})$ is the probability of the user 2D location $\ell_{(x,y)}$ at time $t$, $z$ is AP $a_m$ RSS-Rank and $\textit{dist}(a_m,\ell_{(x,y)})$ is the Haversine distance between AP $a_m$ and 2D-location $\ell_{(x,y)}$.

To verify the Gaussian-based model and estimate its parameters, we conducted an experiment in testbed Bldg. 115 4th floor (Section~\ref{sec:eval}). The RSS values were clustered based on their APs' Haversine distance to the user using hierarchical clustering, yielding the following ranks: very weak (RSS $> -80$), weak ($-70 > $ RSS $ > -80$), mild ($-60 > $ RSS $ > -70$), moderate ($-60 > $ RSS $ > -50$), strong ($-50 > $ RSS $ > -40$) and very strong ($-40 > $ RSS). Each cluster distances were then fitted to a Gaussian distribution after removing the outilers/anomalies~\cite{aly2013dejavu}. We empirically choose 0.8 as the anomalies threshold. Figure~\ref{fig:n_ex} shows examples of the Gaussian distributions from the different RSS-Ranks along with histogram of user's distances from APs with outliers removed for clarity. Note that, \textbf{the RSS-Rank Gaussian model is used to estimate the location PDF in all evaluation testbeds without any training or calibration data collection from the testbeds}.
\subsubsection{Location Estimation}
\vspace*{-1mm}
Using the generated PDF, the user location at time t ($\textit{loc}_t$) is estimated as the weighted centroid of the area where the user is located with probability more than or equal $\ell_\textit{thr}$ (Equation~\ref{eq:loc}). We evaluate the effect of the $\ell_\textit{thr}$ value in Section~\ref{sec:eval}.
\vspace*{-2mm}
\begin{equation}
\textit{loc}_t = \frac{\int_{T} \ell_{(x,y)} \mathcal{L}(p_t^N, \ell_{(x,y)})   \; d\ell_{(x,y)}}{\int_{T} \mathcal{L}(p_t^N, \ell_{(x,y)})   \; d\ell_{(x,y)}}
\label{eq:loc}
\end{equation}
Where $T$ is the area with $\mathcal{L}(p_t^N, \ell_{(x,y)}) \geq \ell_{thr}$.
\subsection{Location Quality Estimation}\label{sec:loc_qual}
Generally, localization systems' accuracy are experimentally validated to %
assess their typical performance under different scenarios using ground-truth data. Although this can help end-users get a sense of the system's expected performance, it fails to give them any clue about the real-time error in their estimated location. Therefore, availability of a quality measure %
 can enhance the system usability for end-users. Moreover, it can help further refine the estimated locations as we show next (Section~\ref{sec:kf}). Typically, the quality measure is represented as a circle of ambiguity with the estimated location at its center and its radius is the predicted quality~\cite{aly2017accurate}.

To predict the quality of the user-location estimate, we analyze the different sources affecting the system's accuracy and model the relation between the accuracy and these sources using regression. Our analysis show that for a $p_t^N$, the following parameters affected the system performance: The visible APs number, the visible APs' average RSS, the maximum RSS and the area of the floor-space with $\mathcal{L}(p_t^N, \ell_{(x,y)}) \geq \ell_{thr}$ , as it represents the uncertainty-level in the user-location PDF. %
Figure~\ref{fig:zeft} shows an example comparing the actual error to the regression model predicted quality. The quality metric has a high precision and recall for instances with high and low accuracy. %
The regression weights are estimated using the same collected data in the 4th floor of testbed Bldg. 115 for the RSS-Rank Gaussian model validation.

\begin{figure*}[!t]
\begin{minipage}{0.23\linewidth}
\center
 \includegraphics[width=\linewidth,height=2.8cm]{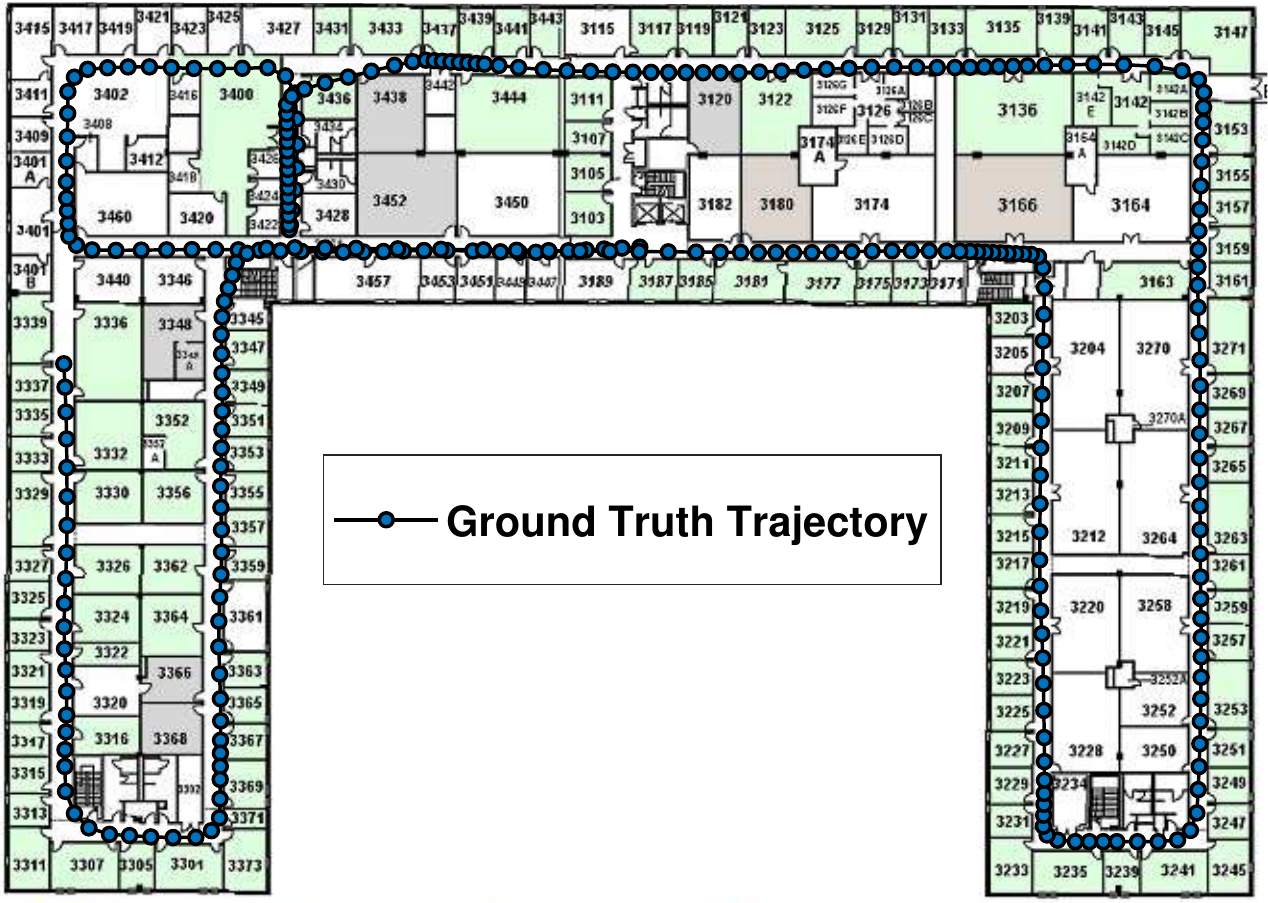}
 \vspace*{-6mm}
\caption{Bldg. 115 3rd floor floorplan with sample trajectories.}
\label{fig:floorplans_ex}
\end{minipage}
\hspace*{1mm}
\begin{minipage}{0.22\linewidth}
\begin{figure}[H]
\center
\includegraphics[width=\linewidth]{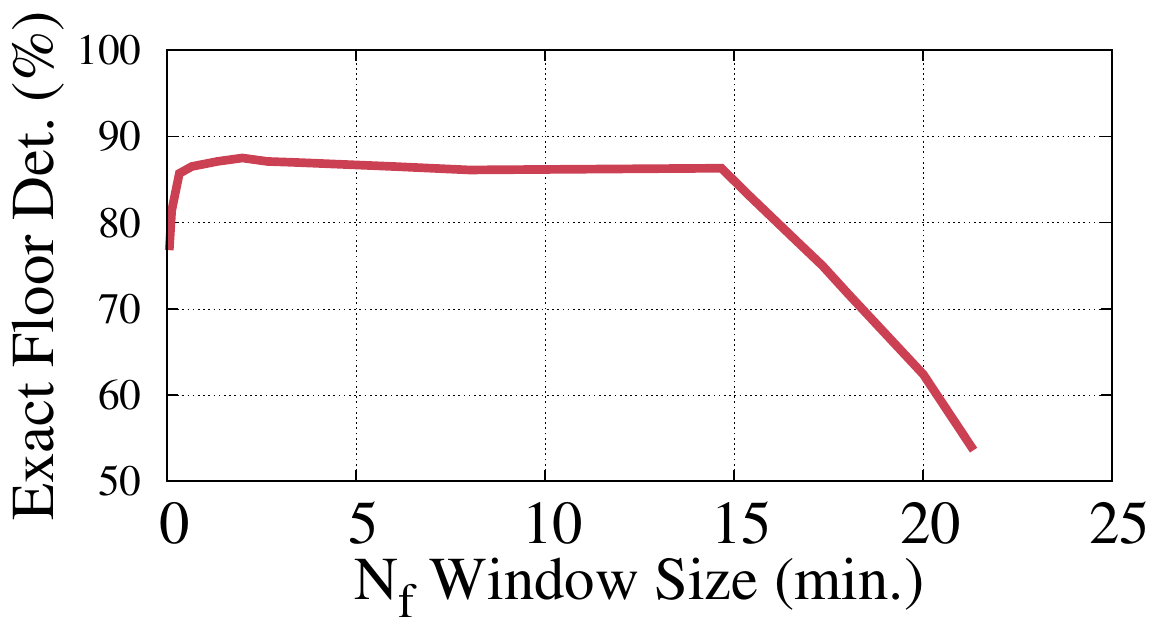}
\vspace*{-4mm}
\caption{Effect of the Floor-level detection $N_f$ value on \sys{}'s exact floor detection accuracy.}
\label{fig:win_nf}
\end{figure}
\end{minipage}
\hspace*{1mm}
\begin{minipage}{0.24\linewidth}
\begin{figure}[H]
\center
\includegraphics[width=0.95\linewidth]{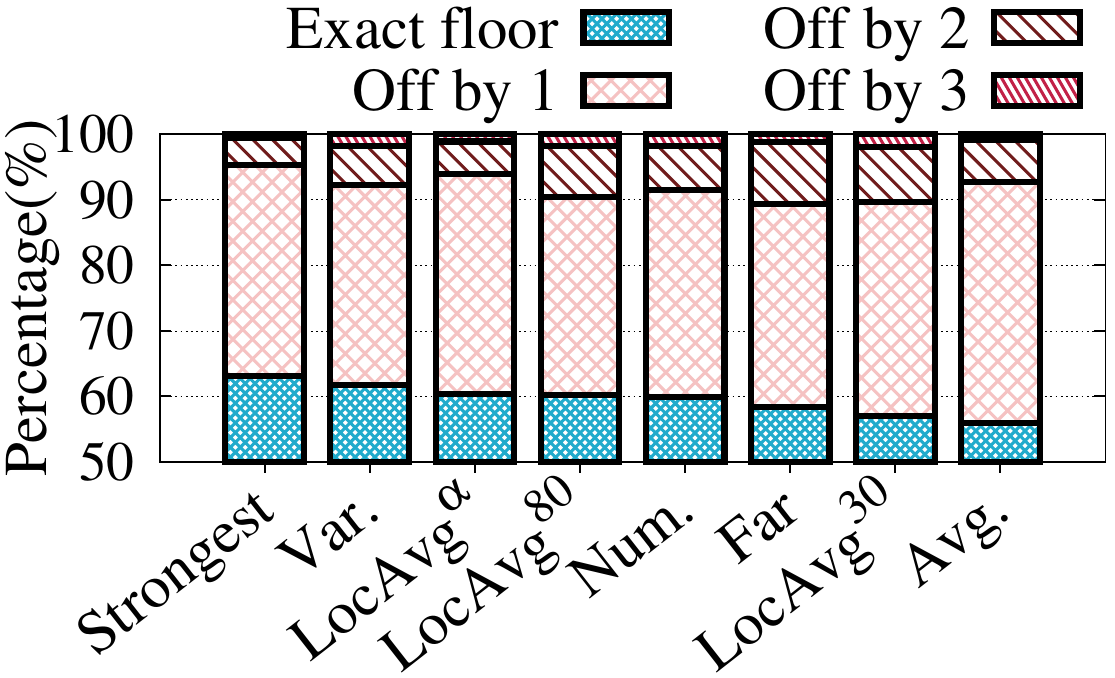}
\vspace*{-4mm}
\caption{Performance of the extracted WiFi features overall the five testbeds.}
\label{fig:fets_perf}
\end{figure}
\end{minipage}
\hspace*{1mm}
\begin{minipage}{0.26\linewidth}
\begin{figure}[H]
\center
\includegraphics[width=0.9\linewidth]{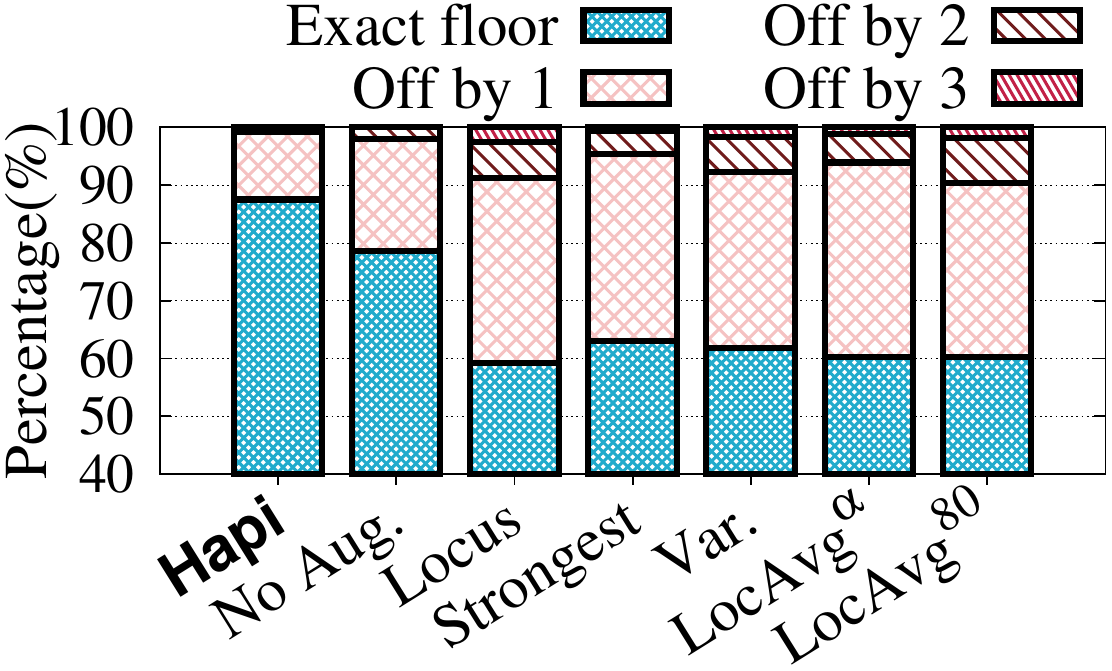}
\vspace*{-4mm}
\caption{Effect of disabling training method on \sys{}'s performance (No Aug. bar) as compared to Locus and the top performing four features.}
\label{fig:training_eff}
\end{figure}
\end{minipage}
\vspace*{-4mm}
\end{figure*}
\subsection{KF Location Refinement}\label{sec:kf}
To refine \sys{}'s location estimate, we use the user history as a window of successive location predictions and their estimated quality through a Kalman Filter (KF)~\cite{grewal2011kalman}. We apply the KF on a window of size $w_k$ and assume that the user is moving with a constant speed over that period. More specifically, we define our state ($\mathbf{s}$) as the user location coordinates ($\mathbf{x}$) and her speed components ($\mathbf{v}$): $\mathbf{s} = [\mathbf{x} \;\; \mathbf{v}]^T$. Thus, the state prediction at time $i$ (prior estimate: $\hat{\mathbf{s}}_i^-$) and its prior variance ($P_i^-$) are estimated as follows:
\begin{equation}
\hat{\textbf{s}}_i^- = A_i \; \hat{\textbf{s}}_{i-1}
\end{equation}
\begin{equation}
P_i^- = A_i \; P_{i-1} \; A_i^T + Q_i
\end{equation}
Where $A_i$ is the state transition matrix at time step $i$, we model it as a constant speed linear motion:
\begin{equation}
A_i =  \begin{bmatrix}
1 & \Delta t_i \\
0 & 1\\
\end{bmatrix}
\vspace*{-2mm}
\end{equation}
$Q_i$ is the process noise at time step $i$ which allows tracking of different forces that could affect the user's movement speed:
\begin{equation}
Q_i =  \begin{bmatrix}
\frac{\Delta t_i^2}{2}\\
\Delta t_i
\end{bmatrix} \times 
\begin{bmatrix}
\sigma_1^2 & 0\\
0  & \sigma_2^2
\end{bmatrix}
\times
\begin{bmatrix}
\frac{\Delta t_i^2}{2}\\
\Delta t_i
\end{bmatrix}^T
\end{equation}
However, since we are doing the KF-refinement over a window rather than continuous tracking, a $\sigma_1 =\sigma_2=0$ gave us good accuracy as the user speed is less likely to change over the window.
Next, using a new measurement, the KF updates the state as follows:
\begin{equation}
K_i = \frac{P^-_i}{P_i^- + R_i}
\end{equation}
\begin{equation}
\hat{\textbf{s}}_i = \hat{\textbf{s}}_{i}^- + K_i (z_i - \hat{\textbf{s}}_{i}^-)
\end{equation}
\begin{equation}
P_i = (1 -K_i)P^-_i
\end{equation}
Where at time step $i$, $K_i$ is the Kalman gain, $R_i$ is the measurement noise and is modeled using the \emph{Location Quality} estimate described in Section~\ref{sec:loc_qual},  $z_i$ is the user estimated location at time $i$ and speed based on her location estimates at times $i$ and $i-1$. Initial state is set as the user location and 0 speed with high variance (100). Empirically, we choose $w_k = 1$min as it balances accuracy refinement with the user location history %
and capturing her motion dynamics.

\begin{table}[!t]
\vspace*{-3mm}
\center
\caption{Summary of the five testbeds and evaluation data.}
\vspace*{-3mm}
\resizebox{1.05\linewidth}{1.3cm}{
\begin{tabular}{|l|c|c|c|c|}\hline
\rowcolor{wolf}
Building & \# Floors & Area (ft$^2$) & \# APs/Floor & \# Samples\\\hline\hline
\rowcolor{aliceblue}
Centreville Hall - \textbf{98} & 9 & 76,340  &[8, 12, 10, 11, 11, 10, 9, 12, 17] & 1837\\
Cumberland Hall - \textbf{122} & 9 & 74,980 & [ 11, 10, 15, 10, 12, 8, 18, 8, 15] & 
2280\\
\rowcolor{aliceblue}
Bel Air Hall - \textbf{99} & 5 & 17,710 & [5, 4, 6, 4, 8] & 223\\
Cambridge Hall - \textbf{96} & 5 & 34,631 & [7, 8, 11, 8, 11] & 1088\\
\rowcolor{aliceblue}
A.V. Williams - \textbf{115}& 4 & 152,130 & [36, 40, 38, 20] & 4995\\\hline\hline
\multicolumn{4}{|c|}{\cellcolor{wolf}Total Number of Samples} & 10423 \\

\hline \end{tabular}
}
\label{tab:testbeds}
\vspace*{-2mm}
\end{table}
\begin{figure*}[!t]
\vspace*{-1mm}
\center
\subfigure[Bldg. = 115]{
 \includegraphics[width=0.1878\linewidth]{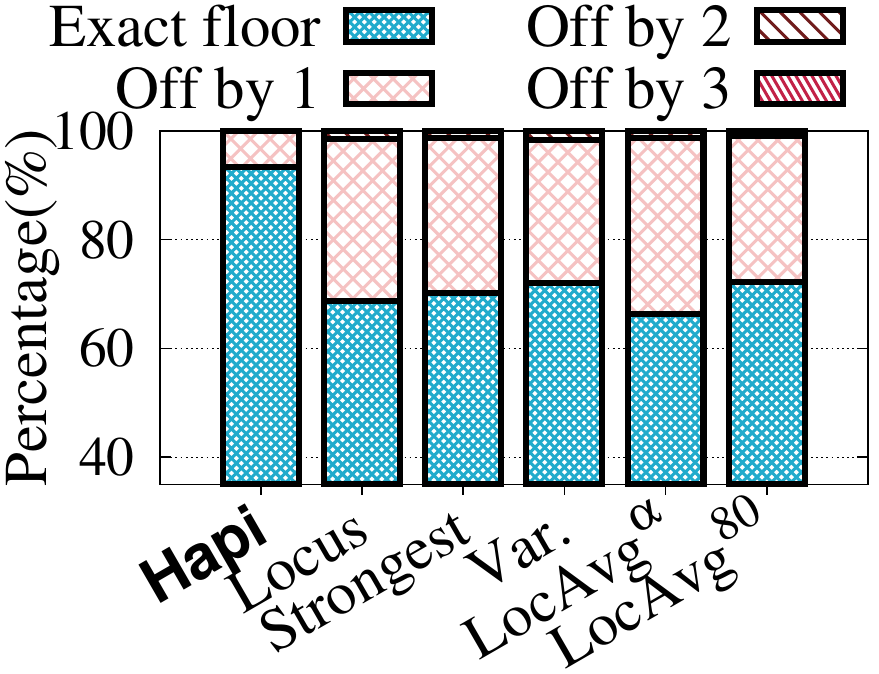}
      \label{fig:acc_bldg_115}
}
\subfigure[Bldg. = 98]{
 \includegraphics[width=0.1878\linewidth]{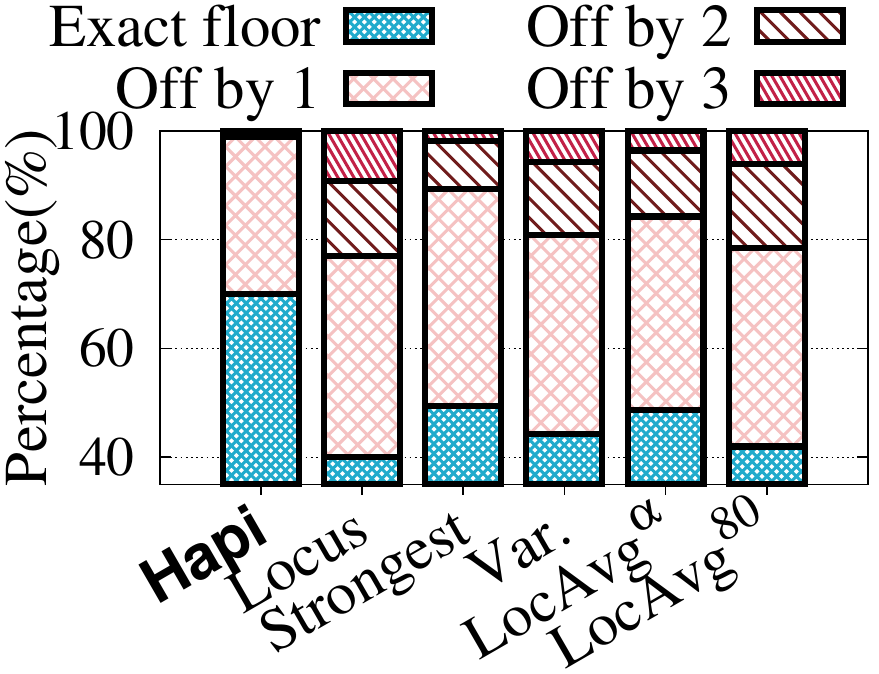}
      \label{fig:acc_bldg_98}
}
\subfigure[Bldg. = 122]{
 \includegraphics[width=0.1877\linewidth]{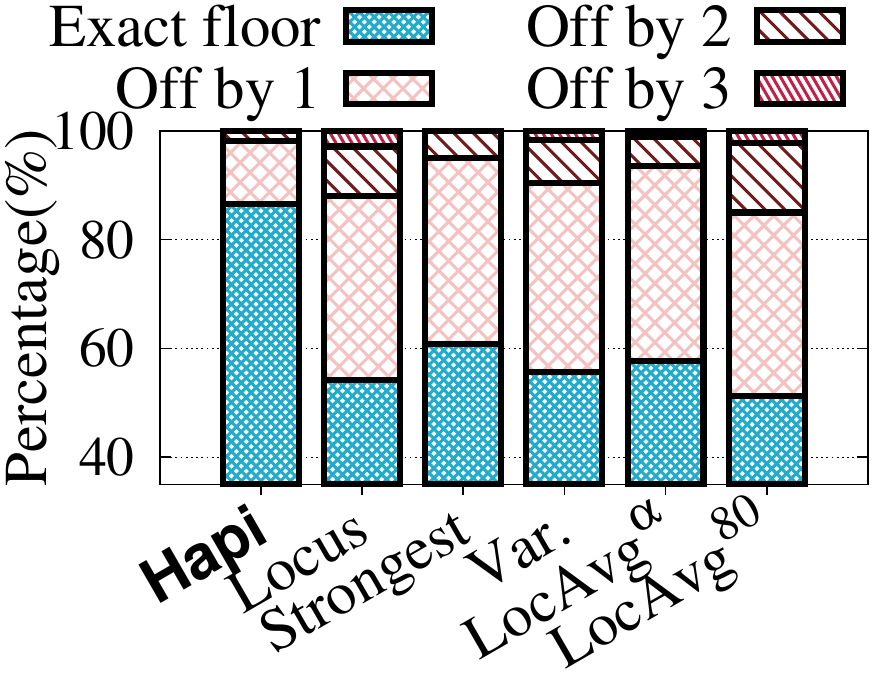}
      \label{fig:acc_bldg_122}
}
\subfigure[Bldg. = 99]{
 \includegraphics[width=0.1877\linewidth]{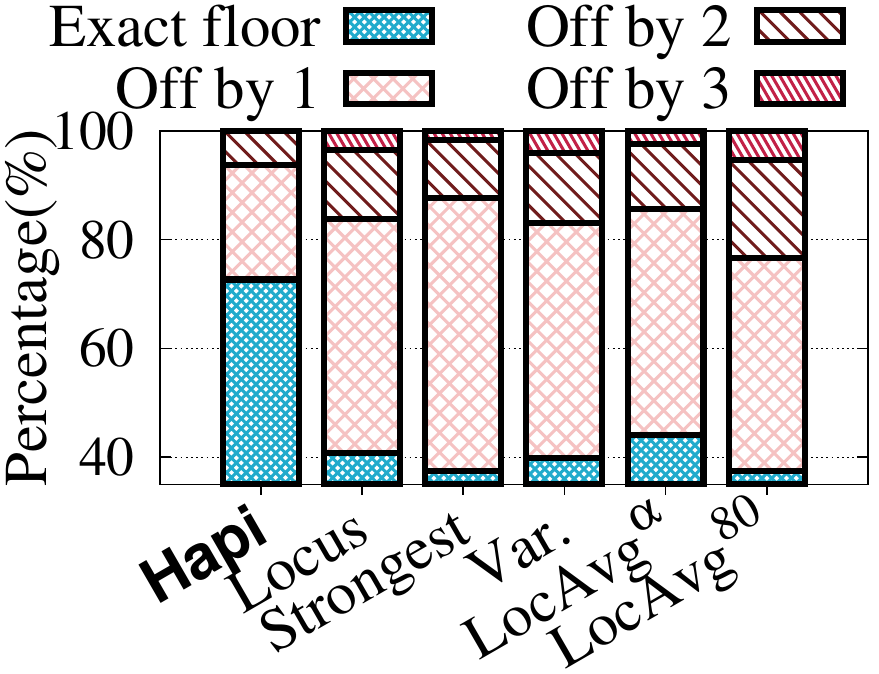}
      \label{fig:acc_bldg_99}
}
\subfigure[Bldg. = 96]{
 \includegraphics[width=0.1877\linewidth]{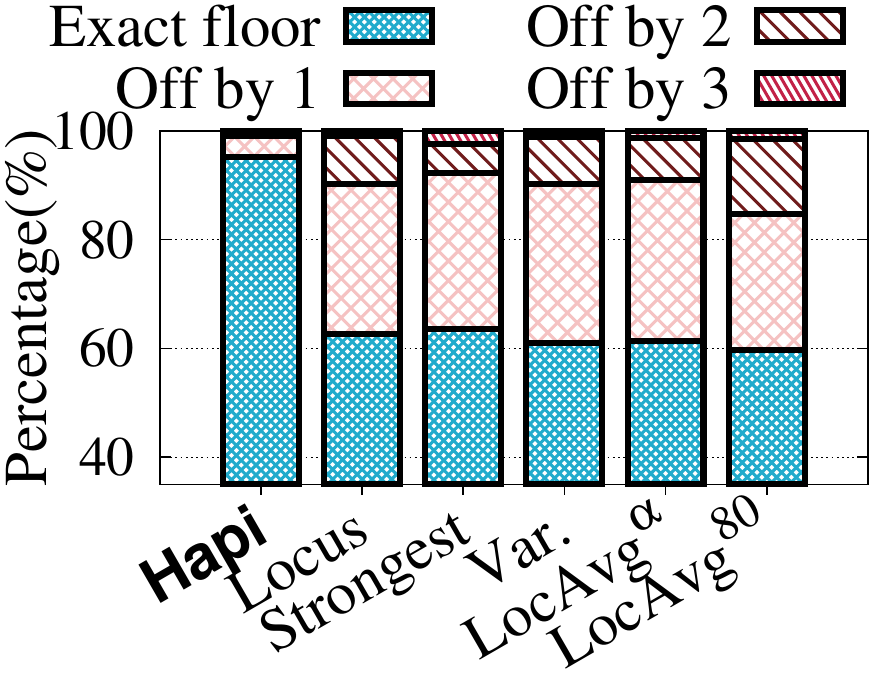}
      \label{fig:acc_bldg_96}
}
\vspace*{-5mm}
\caption{\sys{}'s floor-level detection accuracy in the 5 testbeds as compared to Locus~\protect\cite{bhargava2015locus} and top performing 4 features.}
\label{fig:acc_bldg_2d}
\vspace*{-3mm}
\end{figure*}
\begin{figure*}[!t]
\begin{minipage}{0.68\linewidth}
\begin{table}[H]
\vspace*{-4mm}
\caption{Summary of \sys{}'s accuracy as compared to IncVoronoi~\protect\cite{elbakly2016robust} and Locus~\protect\cite{bhargava2015locus}. Percentages are calculated relative to comparison systems.}
\vspace*{-3mm}
\resizebox{\linewidth}{1.2cm}{
\begin{tabular}{|l|c|c|c||c|c|c|}\hline
\rowcolor{almond}
\multirow{1}{*}{Testbed} & \multicolumn{3}{c||}{Exact Floor-Level Detection Percentage} & \multicolumn{3}{c|}{2D Location Median Error}\\\cline{2-7}
\cellcolor{almond}  & \sys{} &  IncVoronoi~\cite{elbakly2016robust} & Locus~\cite{bhargava2015locus} & \sys{} & IncVoronoi~\cite{elbakly2016robust} & Locus~\cite{bhargava2015locus} \\\hline\hline
\rowcolor{azure}
Bldg. 99 & \textbf{72.6\%} & N/A & 40.7\% \emph{(+78.4\%)} & \textbf{3.1m} & 5.1m \emph{(+39.2\%)}& 23.5m \emph{(+86.8\%)}\\\hline
Bldg. 96 & \textbf{95.2\%} & N/A & 62.6\% \emph{(+52.1\%)}& \textbf{3.4m} &6.0m \emph{(+43.3\%)}& 13.9m  \emph{(+75.5\%)}\\\hline
\rowcolor{azure}
Bldg. 122 & \textbf{86.5\%} & N/A & 54.2\% \emph{(+59.6\%)}& \textbf{3.5m} & 6.2m  \emph{(+43.5\%)} & 15.3m  \emph{(+77.1\%)}\\\hline
Bldg. 115 & \textbf{93.4\%} & N/A & 68.7\% \emph{(+36.0\%)}&  \textbf{3.6m} & 11.8m  \emph{(+69.5\%)}& 13.0m  \emph{(+72.3\%)}\\\hline
\rowcolor{azure}
Bldg. 98 & \textbf{70\%} & N/A & 40.1\% \emph{(+74.6\%)} & \textbf{3.5m} &6.5m  \emph{(+46.2\%)}& 17.9m  \emph{(+80.4\%)}\\\hline\hline
\cellcolor{almond} Overall &\textbf{87.5\%} & N/A & 59.2\% \emph{(+47.7\%)} & \textbf{3.5m} & 8.9m  \emph{(+60.7\%)}& 14.6m  \emph{(+76.0\%)}\\\hline
\end{tabular}
}
\label{tab:acc_sum}
\end{table}
\end{minipage}
\hspace*{1mm}
\begin{minipage}{0.3\linewidth}
\center
{
\includegraphics[width=0.9\linewidth]{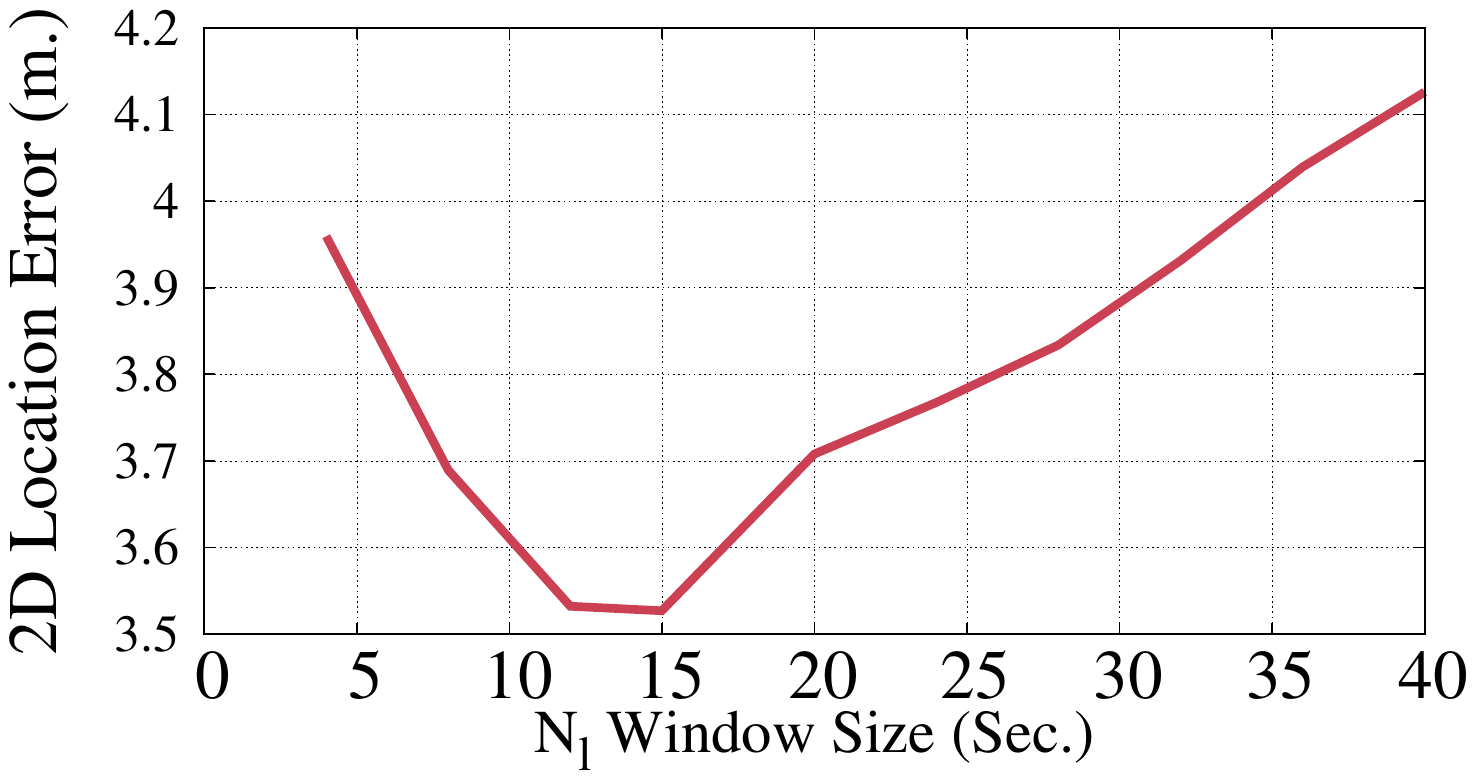}
\vspace*{-3mm}
\caption{Effect of the $N_l$ value on \sys{}'s 2D location accuracy.}
\label{fig:nl_eval}
}
\end{minipage}
\vspace*{-3mm}
\end{figure*}

\section{Evaluation}\label{sec:eval}
We implemented \sys{} using a client-server architecture where the client is an Android mobile app that collects the raw WiFi-scans through the Android API and sends them to the server. The localization algorithm is implemented as a web-service on the server. To evaluate the system accuracy under different WiFi-network characteristics, we used \emph{five different testbeds} that exhibit different floor-plans, building structures and number of floors. \emph{A total of 10423 samples were collected by 13 subjects (7 females and 6 males) within a period of 6 months}. The subjects used different android devices including LG Nexus 5X, LG Nexus 5, Motorola Droid Razr HD, Moto G4, Samsung Galaxy Tab 4, Samsung Galaxy J7, Samsung Galaxy J3, Samsung Galaxy S III, OnePlus 3 and Huawei Mate 9. To obtain the ground truth, we developed a special app using Google Indoor Maps~\cite{gindoormaps}, where subjects report their floor-level ground truth. Afterwards, the app shows them their current floorplan indoor maps and they mark their ground-truth 2D locations on the map as they walk. Thus, for every WiFi-scan the app reports a pseudo-3D ground-truth location. Figure~\ref{fig:floorplans_ex} shows Bldg. 115 (one of the 5 testbeds) 3rd floor floorplan and sample ground-truth user trajectories. The full-list of the testbeds' floorplans can be found here~\cite{avwfloorplans,Cambridgefloorplans,BelAirfloorplans,Cumberlandfloorplans,centrevillefloorplans} and Table~\ref{tab:testbeds} summarizes the testbeds characteristics.

For the rest of this section, we start by evaluating \sys{}'s floor-level detection performance. Then, we evaluate its 2D location accuracy. In addition, we compare its accuracy to two state-of-the-art calibration-free systems: Locus~\cite{bhargava2015locus} and IncVoronoi~\cite{elbakly2016robust}.
\subsection{Floor-Level Detection Accuracy}
We start by evaluating the effect of \sys{}'s floor-level detection module's different parameters and components. Then, we show \sys{}'s performance in the 5 testbeds as compared to Locus~\citep{bhargava2015locus}. \emph{Note that, we do not compare to IncVoronoi~\cite{elbakly2016robust} here as it assumes a single-floor area of interest and estimates the user 2D location only.}
\subsubsection{Effect of the Profile Window Size}\label{sec:eval_nf}
Figure~\ref{fig:win_nf} shows the effect of the profile window size $N_f$ on \sys{}'s
exact floor detection accuracy. The WiFi profile helps better capture the user surrounding WiFi network. Hence, increasing $N_f$ increases the system accuracy. However, as the window gets too large, the accuracy decreases, as the user can move to another floor. We set $N_f$ to $2$mins. as it balances accuracy and being a reasonable duration for the user dwelling period in a single floor. %
\subsubsection{Performance of the Different WiFi Features}
Figure~\ref{fig:fets_perf} shows the overall performance of the different WiFi features for the 5 testbeds. %
Using each feature, we set the user-floor as the floor with the maximum feature value. For example, for the ``Number of APs'' feature (\textit{Num}), the user-floor becomes the floor with the maximum number of visible APs.  Generally, we can identify the user floor around 60\% of the time using any of the features. However, as you can see in Figure~\ref{fig:acc_bldg_2d}, each performed differently in the different testbeds due to their varied WiFi network and building characteristics. Interestingly, using the APs' pseudo-3D location through the proposed \emph{Local Average Signal Strength} ($LocAvg$) feature performed better than using the visible APs' average signal strength (\textit{Avg}). In addition, selecting the proximity region using $\alpha$ performed better than other values because it varies based on the user profile.%
\subsubsection{Effect of our DL Training Methodology}
Figure~\ref{fig:training_eff} compares \sys{}'s floor-level detection to same DNN architecture without the data generalization and balancing during training (\emph{No Aug.} bar on figure), Locus and the top performing four features. We can see that the DNN-method improves the accuracy over Locus and the individual techniques. Moreover, our proposed data generalization and balancing lead to an improvement of 11.3\% in the user exact floor detection.
\subsubsection{Comparison with Other Systems}
Figure~\ref{fig:acc_bldg_2d} shows \sys{}'s floor-level detection accuracy as compared to Locus~\cite{bhargava2015locus} and top performing WiFi features in the 5 testbeds. Table~\ref{tab:acc_sum} summarizes the results. \sys{}'s deep-learning model takes the building's network architecture into account and yields a better performance compared to individual features achieving up to $95.2\%$ exact floor detection in Bldg. 96 with no training data from any of the buildings. In addition, this is better than Locus~\cite{bhargava2015locus} by $52.1\%$. Locus~\cite{bhargava2015locus} algorithm was tuned using heuristics from experimentation in one building, this lead to decrease in its accuracy when applied in larger-scale. Also, Locus uses only the APs' installation floor.%
\begin{figure*}[!t]
\vspace*{-1mm}
\begin{minipage}{0.25\linewidth}
\center
\includegraphics[width=0.95\linewidth]{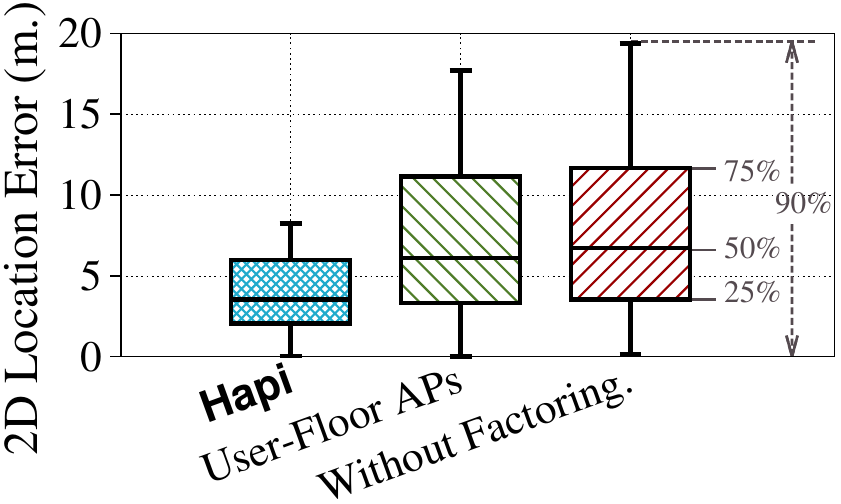}
\vspace*{-4mm}
\caption{Effect of disabling \sys{}'s Floor Attenuation Factoring module and using APs from the user floor only.}
\label{fig:perf_floor}
\end{minipage}
\hspace*{0.5mm}
\begin{minipage}{0.23\linewidth}
\center
\includegraphics[width=0.95\linewidth]{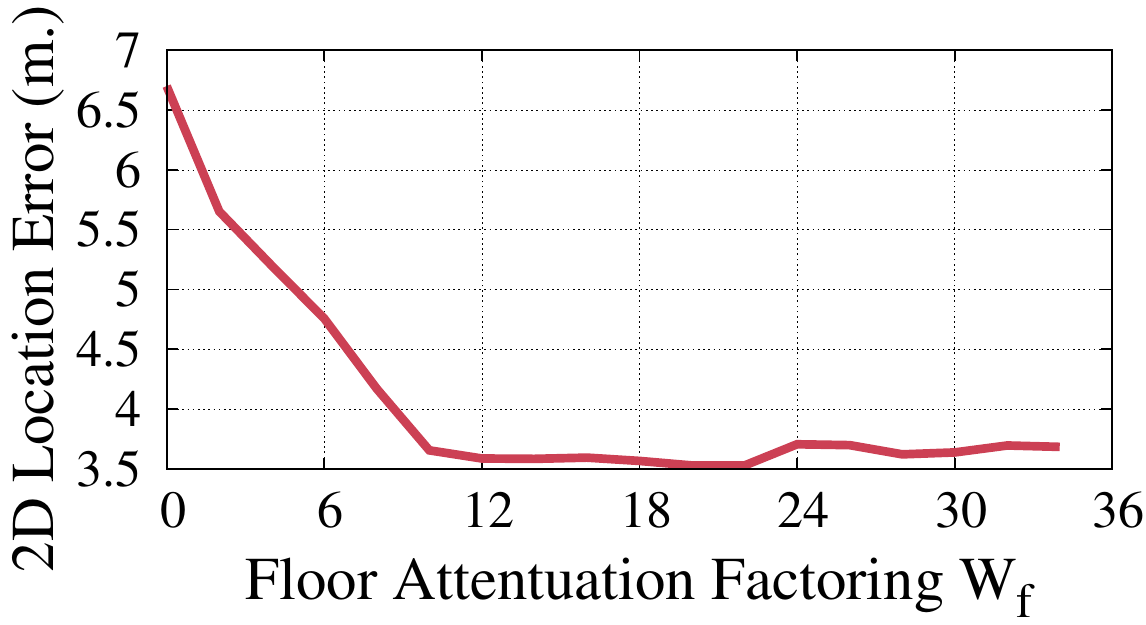}
\vspace*{-2mm}
\caption{Effect of the $W_f$ value on \sys{}'s 2D location accuracy.}
\label{fig:faf_w}
\end{minipage}
\hspace*{0.5mm}
\begin{minipage}{0.24\linewidth}
\center
\includegraphics[width=0.95\linewidth]{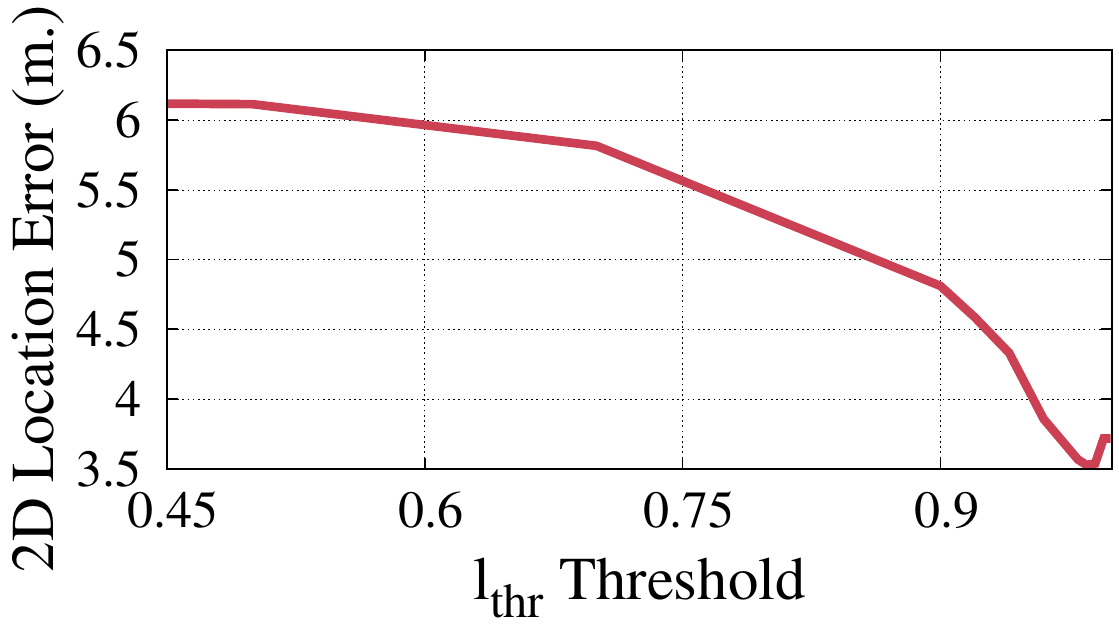}
\vspace*{-3mm}
\caption{Effect of the location estimation $\ell_{thr}$ on \sys{}'s 2D location accuracy.}
\label{fig:l_thr}
\end{minipage}
\hspace*{0.5mm}
\begin{minipage}{0.24\linewidth}
\center
\includegraphics[width=0.95\linewidth]{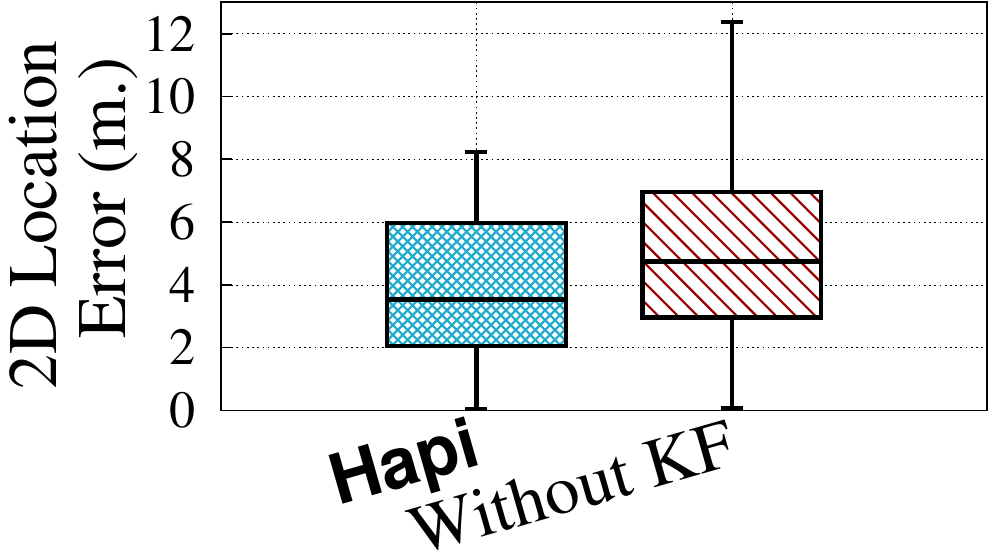}
\vspace*{-3mm}
\caption{Effect of the Kalman Filter module on \sys{}'s 2D location accuracy.}
\label{fig:perf_kf}
\end{minipage}
\vspace*{-7mm}
\end{figure*}
\begin{figure*}[!t]
\center
\subfigure[Bldg. 115]{
 \includegraphics[width=0.18\linewidth,height=3cm]{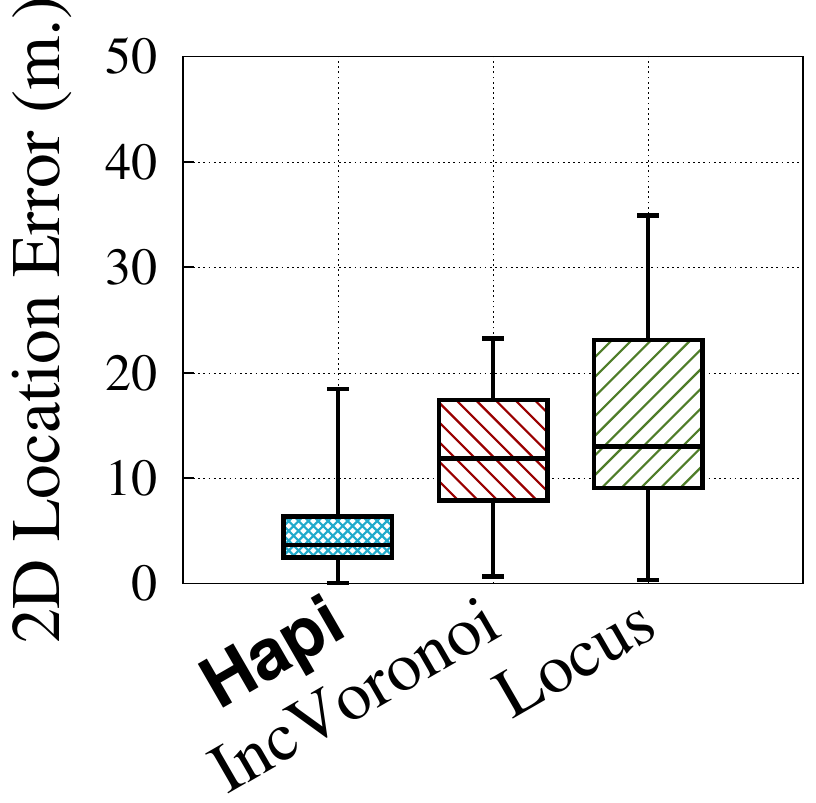}
      \label{fig:acc2_bldg_115}
}
\subfigure[Bldg. 98]{
 \includegraphics[width=0.18\linewidth,height=3cm]{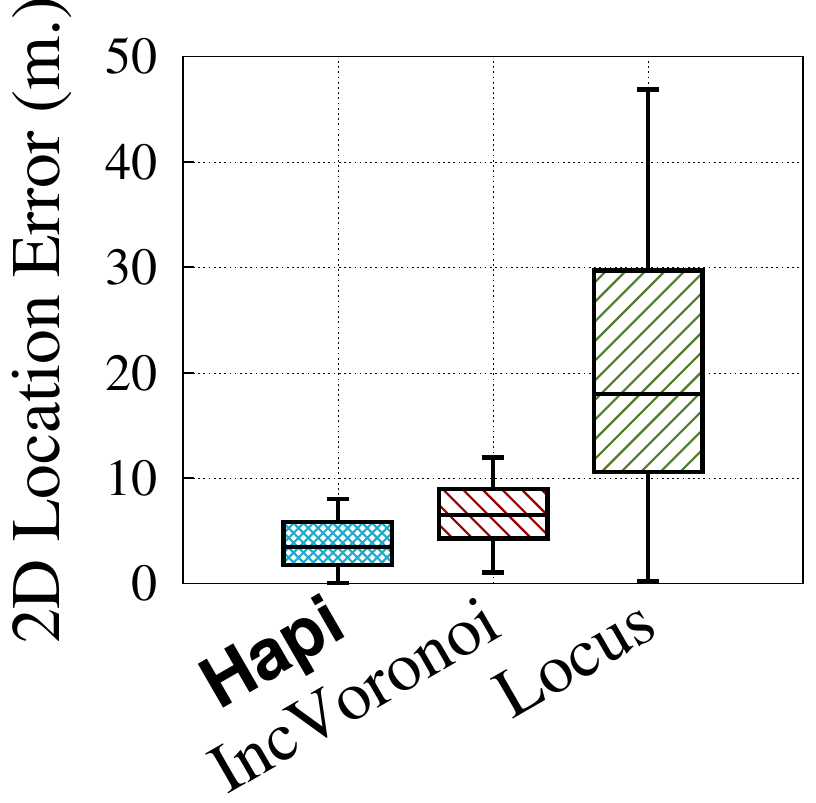}
      \label{fig:acc2_bldg_98}
}
\subfigure[Bldg. 122]{
 \includegraphics[width=0.18\linewidth,height=3cm]{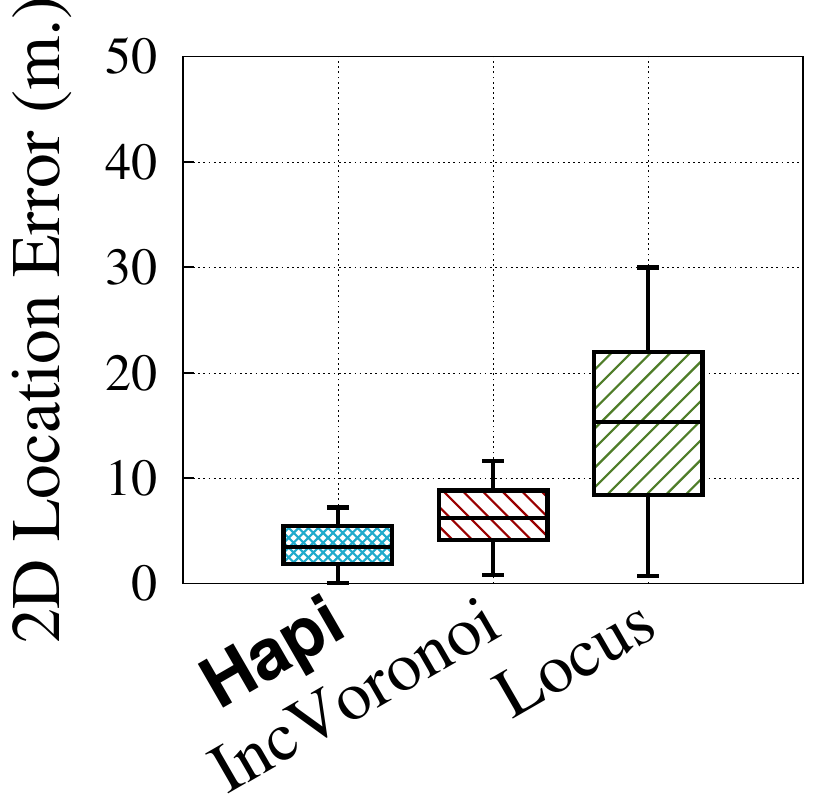}
      \label{fig:acc2_bldg_122}
}
\subfigure[Bldg. 99]{
 \includegraphics[width=0.18\linewidth,height=3cm]{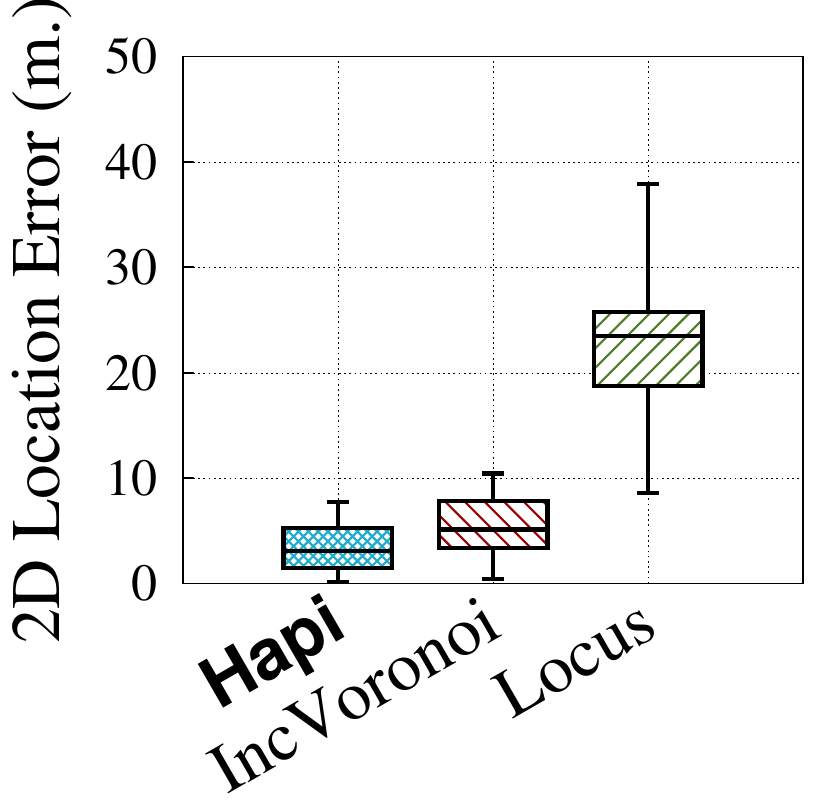}
      \label{fig:acc2_bldg_99}
}
\subfigure[Bldg. 96]{
 \includegraphics[width=0.18\linewidth,height=3cm]{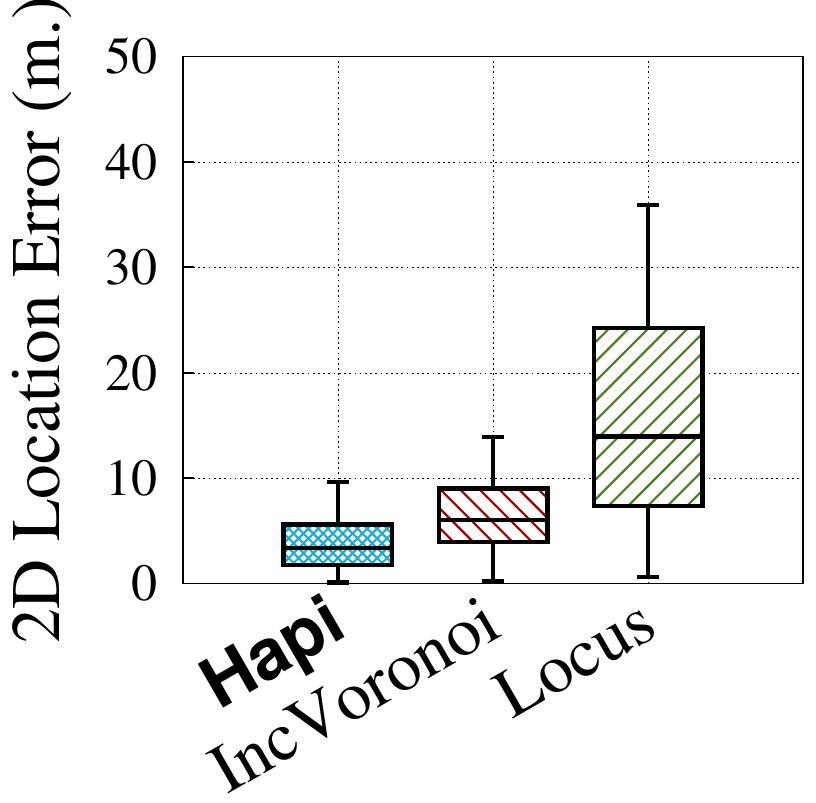}
      \label{fig:acc2_bldg_96}
}
\vspace*{-5mm}
\caption{\sys{}'s 2D location accuracy in the 5 testbeds as compared to IncVoronoi~\protect\cite{elbakly2016robust} and Locus~\protect\cite{bhargava2015locus}.}
\label{fig:acc2_bldg}
\vspace*{-3mm}
\end{figure*}
\vspace*{-3mm}
\subsection{2D Localization Accuracy}
We start by evaluating the effect of \sys{}'s 2D location estimation algorithm different parameters and components. Then, we show \sys{}'s performance in the 5 testbeds as compared to IncVoronoi~\cite{elbakly2016robust} and Locus~\citep{bhargava2015locus}.
\subsubsection{Effect of the Profile Window Size}
Figure~\ref{fig:nl_eval} shows the effect of the profile's $N_l$ value on the user 2D location estimation. The WiFi profile helps better capture the user surrounding WiFi network and thus a larger $N_l$ leads to a better accuracy. %
However, as $N_l$ increases, the user may have moved away from her location leading to accuracy reduction. %
 Thus, we choose $N_l=15$ seconds. Note that, as expected, the $N_l$ value is shorter than the $N_f$ value (Section~\ref{sec:eval_nf}), as users are more likely to stay longer on the same floor as opposed to their location within the floor.
\subsubsection{Effect of the Floor Attenuation Factoring}
Figure~\ref{fig:perf_floor} shows the effect of the \emph{Floor Attenuation Factoring} module on \sys{}'s 2D location accuracy. The figure emphasizes the importance of using visible APs from all floors while applying the Floor Attenuation Factoring. We can see that it improves the accuracy over just using APs from the user floor by 42.6\% and over using the APs without the Floor Attenuation Factoring by 47.8\%. Note that using APs from other floors without processing their RSS leads to a decrease in the accuracy as they represent noisy/misleading measurements added to the user location PDF as discussed in Section~\ref{sec:faf}.

Figure~\ref{fig:faf_w} shows the effect of the Floor Attenuation Factoring module $W_f$ parameter on the location accuracy. Increasing $W_f$ improves the accuracy as it compensates for the AP's RSS attenuation due to its installation floor. Yet, as it increases, it can end up mapping all RSS values from APs in other floors to a Very Strong Rank which causes a decrease in accuracy. We set $W_f$ to 15 to balance the RSS mapping as seen in the figure.
\subsubsection{Effect of the Location Estimation Threshold}
Figure~\ref{fig:l_thr} shows the effect of the location estimation threshold ($\ell_\textit{thr}$) on \sys{}'s location accuracy. Increasing $\ell_\textit{thr}$ leads to removing uncertain areas when calculating the user location and consequently improving the system's location accuracy. However, as $\ell_\textit{thr}$ increases, it may lead to removing areas where the user can be located and decreasing the location accuracy. %
Thus, we set $\ell_\textit{thr}$ to $0.98$.
\subsubsection{Effect of the KF Refinement}
Figure~\ref{fig:perf_kf} shows the effect of the KF Location Refinement on \sys{}'s 2D location accuracy. The KF reduces the maximum errors significantly and the median error by 25.5\%, as it considers both the user's motion history and the quality of the location instances.
\subsubsection{Comparison with Other Systems}
Figure~\ref{fig:acc2_bldg} shows \sys{}'s 2D location accuracy in the 5 testbeds and compares it to IncVoronoi~\cite{elbakly2016robust} and Locus~\cite{bhargava2015locus}, and Table~\ref{tab:acc_sum} summarizes the results. \sys{} exhibited a better accuracy in all testbeds. We can see that Locus using, only, the user-floor's APs' raw RSS values through triangulation resulted in a lower accuracy when compared to \sys{} and IncVoronoi~\cite{elbakly2016robust} relative RSS approaches. Specifically, \sys{} had 76.0\% improvement in median accuracy over Locus overall testbeds. While IncVoronoi had better performance than Locus, using the user floor to preprocess the RSS and our proposed novel RSS-Rank Gaussian model enabled \sys{} to improve the median accuracy consistently and achieving an improvement of 60.7\% overall testbeds. %
\section{Related Work}\label{sec:relatedWork}
WiFi presents an attractive indoor localization technology due to its world-wide ubiquity. However, due to the wireless signal propagation sensitivity to noise in typical multipath-rich indoor environments~\cite{hossain2015survey,yang2015mobility}, researchers proposed to construct an offline fingerprint map for the area of interest and use it in real-time to localize users, e.g.~\cite{youssef2005horus,honkavirta2009comparative}. Nevertheless, conducting site-surveys, to construct and maintain the fingerprint map for every deployment area, increases the system's deployment overhead. Crowd-sourcing based systems were introduced to reduce this overhead along with fusing WiFi with other sensors such as camera, accelerometer, gyroscope, etc...~\cite{yang2015mobility}. For example, in~\cite{wang2012no,abdelnasser2016semanticslam} authors crowd-source the smartphone's WiFi and inertial sensors to build a map of landmarks including stairs, spots with unique magnetic fluctuations or corners with unique WiFi signatures to localize users. In~\cite{dong2015imoon}, the iMoon system crowdsources 2D photos to build 3D models of the environment and uses WiFi fingerprints to speed its image-based localization. Using additional sensors can limit the ubiquity and increase the power consumption of the system. Moreover, crowdsourcing brings its own challenges including privacy concerns and user participation incentives, among others. Automatic theoretical-based map construction methods were also presented, e.g. AROMA~\cite{eleryan2011aroma,aly2014analysis}.
These systems rely on the APs' RSS absolute value for localization. In~\cite{machaj2011rank,liu2016selective,elbakly2016robust}, authors propose to use the APs relative order instead which \emph{improves the localization system resilience to heterogeneity}. In~\cite{machaj2011rank} authors proposed to fingerprint the APs' RSS rank. This approach was extended in~\cite{liu2016selective} and IncVoronoi~\cite{elbakly2016robust} to remove the fingerprint map calibration requirements. In~\cite{liu2016selective}, authors assume a universal propagation model of the APs' signals and use that to construct a map of the APs' sequence based on their predicted RSS value. Then, in real-time, the user is located based on her visible APs' sequence similarity. Alternatively, instead of ordering the visible APs sequence, IncVoronoi~\cite{elbakly2016robust} employs a pairwise comparison of the user's visible APs to identify her most probable location through Voronoi tessellation. 
Similarly, \sys{} is calibration-free and refrains from using the APs' RSS absolute value. However, instead of solely relying on the APs order which totally ignores the APs' RSS levels, we propose an RSS-Rank Gaussian-based model that balances heterogeneity resilience and accuracy. Moreover, \sys{} uses APs from all floors to identify the user pseudo-3D location, allowing it to work in realistic multistory buildings.

Recently, WiFi-based localization systems have been proposed to identify the user's pseudo-3D location. In~\cite{kim2018scalable}, authors propose to fingerprint the entire building. However, this increases the deployment and maintenance overhead significantly, particularly in buildings with a large number of floor-levels. ViFi~\cite{caso2016vifi} uses the Multi-Wall Multi-Floor propagation model~\cite{action1999231} to automatically predict the APs' signal propagation and construct the multifloor fingerprint maps. However, theoretical models accuracy degrades significantly in uncontrolled environments~\cite{caso2016vifi,elbakly2016robust}. 
Contrarily, Locus~\cite{bhargava2012locus,bhargava2015locus} is calibration-free and uses a  heuristics-based algorithm to identify the user's floor. Then, employs weighted triangulation for \emph{APs from that floor only} to estimate her 2D location. Yet, this trades accuracy for ease of deployment as we have quantified in Section~\ref{sec:eval}.  TrueStory~\cite{truestory} is also a calibration-free system that estimates the user floor using a Multilayer Perceptron. However, it only uses the APs' floor number and ignores its location within the floor. As we have shown in Section~\ref{sec:eval}, this can limit the system accuracy. Additionally, these systems were tested in up to 4-5 floors buildings only. Similarly, \sys{} is calibration-free. However, it identifies the user's floor using a scalable and general DL-based method and employs an RSS-Rank probabilistic model to identify her location on that floor accurately and robustly using \emph{APs from all floors}. In addition, we identify \sys{}'s location estimates quality and use it within a KF to further refine the accuracy. The system was thoroughly tested in up to 9 floors multistory buildings.

There have been some recent efforts to boost WiFi-based indoor localization systems' accuracy using Channel State Information (CSI) instead of the APs' RSS~\cite{yang2013rssi}. For example, the SpotFi~\cite{kotaru2015spotfi} and the PhaseFi~\cite{wang2016csi} systems which employed angle of arrival and fingerprinting based approaches respectively. Both systems were tested using the Intel 5300 WiFi NIC in a single-floor testbed area.
  CSI-based methods are not suitable for smartphone-based localization due to the unavailability of the physical layer information from the smartphones' NICs and the mobile operating systems' APIs. On the other hand, \sys{} leverages the widely available AP's RSS information to provide an accurate and robust pseudo-3D location in multistory buildings.

\section{Conclusion}\label{sec:conc}
We presented the design, implementation and evaluation of
\sys{}, a novel calibration-free WiFi-based indoor localization
system that works in realistic multistory deployment environments. It identifies the user's pseudo-3D location: her floor-level and 2D-location on that floor. We described \sys{}'s basic idea and showed how it combines a deep-learning based method to identify the user's floor, with an RSS-Rank Gaussian-based method to estimate the user's 2D location on that floor. Moreover, we present a regression-based method to predict \sys{}'s location estimates' quality and employ it within a KF to further refine the location accuracy.  %

Implementation of \sys{} on a wide-range of android devices, with 13 subjects over 6 months in 5 different up to 9 floors multistory buildings, show that \sys{} can identify the user's exact floor upto 95.2\% of the time and her 2D location on that floor with a median accuracy of 3.5m, achieving an improvement of 52.1\% and 76.0\% respectively over related calibration-free state-of-the-art systems.

Currently, we are expanding \sys{} in multiple directions including implementing the algorithm from the APs-side, using unsupervised WiFi-scans to improve the accuracy and combining it with indoor map-matching, among others.

\section*{Acknowledgment}
This research was supported in part by a Google Scholarship and the Prometheus-UMD which was sponsored by the DARPA BTO under the auspices of Col. Matthew Hepburn through agreement [N66001-17-2-4023 and/or N66001-18-2-4015]. The findings and conclusions in this report are those of the authors and do not necessarily represent the official position or policy of the funding agency and no official endorsement should be inferred. %
\begin{spacing}{0.5}
\bibliographystyle{ACM-Reference-Format}
\bibliography{hapi} 
\end{spacing}
\end{document}